\documentclass[twocolumn,amsmath,superscriptaddress,amssymb,nopacs,nofootinbib,prx]{revtex4-1}

\usepackage[usenames,dvips]{color}
\usepackage{graphicx}
\usepackage{dcolumn}
\usepackage{bm}
\usepackage{mathtools}
\usepackage{epstopdf}
\usepackage{esint}
\usepackage{xcolor}
\usepackage{dsfont}
\usepackage{amsmath}
\usepackage{nicefrac}

\usepackage[colorlinks=true,citecolor=blue,linkcolor=magenta, urlcolor=magenta]{hyperref}

\newcommand{\bea}{\begin{eqnarray}}
\newcommand{\eea}{\end{eqnarray}}
\newcommand{\bt}{\textbf}

\newcommand{\phd}{\phantom{\dag}}
\newcommand{\ph}{\phantom{.}}

\newcommand{\noi}{\noindent}
\newcommand{\no}{\nonumber}

\begin{document}
\def\v#1{{\bf #1}}

\title{Majorana Braiding Racetracks from\\ Charge Chern Insulator - Superconductor Hybrids}

\author{Jun-Ang Wang}
\affiliation{CAS Key Laboratory of Theoretical Physics, Institute of Theoretical Physics, Chinese Academy of Sciences, Beijing 100190, China}
\affiliation{School of Physical Sciences, University of Chinese Academy of Sciences, Beijing 100049, China}
\author{Sen Zhou}
\affiliation{CAS Key Laboratory of Theoretical Physics, Institute of Theoretical Physics, Chinese Academy of Sciences, Beijing 100190, China}
\affiliation{School of Physical Sciences, University of Chinese Academy of Sciences, Beijing 100049, China}
\affiliation{CAS Center for Excellence in Topological Quantum Computation, University of Chinese Academy of Sciences, Beijing 100049, China}
\author{Panagiotis Kotetes}
\email{kotetes@itp.ac.cn}
\affiliation{CAS Key Laboratory of Theoretical Physics, Institute of Theoretical Physics, Chinese Academy of Sciences, Beijing 100190, China}

\vskip 1cm
\begin{abstract}
Recent experiments have provided evidence for chiral charge order in Kagome superconductors (SCs). This intriguing possibility motivates us to unveil the first pathway to engineer topological superconductivity by harnessing the interplay of charge Chern insulators (CIs) and conventional SCs. We here identify under which conditions a pyramidal SC/CI/SC he\-te\-ro\-structure induces an effective 1D spinless p-wave SC that allows pinning Majorana zero modes (MZMs) at termination edges and domain walls. As we reveal, such a MZM track is controlled by the phase difference of the two SCs involved and additional magnetic fields, which are required for ge\-ne\-ra\-ting Rashba-like spin-orbit coupling. Further, we show that a SC/CI/SC/CI/SC double-pyramidal hybrid defines a double MZM track, in which braiding occurs by varying the two superconducting phase differences in space and adiabatically in time. Given the geometry of the MZM racetrack, we propose to employ the time-averaged quadrupolar dif\-fe\-ren\-tial conductance to confirm the here-termed MZM track exchange process which is pivotal for braiding. In addition, we identify experimental knobs which enable the fusion of MZM pairs, and the detection of the underlying non-Abelian to\-po\-lo\-gi\-cal order and twofold many-body ground state degeneracy by encoding it in a topological invariant.      
\end{abstract}

\maketitle

\section{Introduction}

Groundbreaking experiments have recently suggested the emergence of chiral charge order in the family of Kagome superconductors (SCs) AV$_3$Sb$_5$, with A=K~\cite{UCDW_KVSb,muonGraf,muonKVSb,OpticalDetecCDW,NematicCdwKVSb,XrayEPcoupling}, A=Cs~\cite{AHE_CDW_CVSb,CascadeofPhasesCVSb,CDWvortexCVSb,FSmappingCVSb,HiddenFluxPhaseCVSb}, and~A=Rb~\cite{UCDW_RbVSb}. Up to date, the above experiments and related theoretical works~\cite{BinghaiCDW,ChiralFluxKagome,Titus,Nandkishore,KunJiangClassification,BinghaiCDWGeometry,CDWChristensen} indicate that a specific type of triple-Q charge order~\cite{Venderbos} appears, and opens a gap at the M point of the Brillouin zone. The emergence of the charge order induces ground state loop currents, and appears to be driven by a phonon instability~\cite{XrayEPcoupling} which is stabilized by virtue of van Hove singularities~\cite{KagomeRichInvHS} in the electronic density of states. In its insulating regime, this state of matter has a topological character and is predicted to feature a nonzero Chern number~\cite{ChiralFluxKagome,Nandkishore}, which {\color{black}can in turn lead} to electronic chiral edge modes for strip sample geometries. Assuming the absence of an odd-under-inversion spin-orbit coupling (SOC), these edge modes are spin degenerate. Therefore, the chiral charge order in these Kagome SCs becomes topologically equivalent to the charge Chern insulator (CI) that was originally proposed by Haldane~\cite{Haldane}. 

In spite of its long\-stan\-ding history, the charge CI is still considered to be an elusive state of matter, with the high-$T_c$ cuprates having so far constituted the most prominent material candidates for its realization. In the cuprates, a number of theoretical groups have proposed va\-rious interaction-driven loop-current orders. These theories are divided into two categories, i.e., either translationally-invariant orders driven by multi-orbital effects, which were put forward by Varma~\cite{VarmaPRB97,VarmaPseudogap,VarmaJCMP}, or, order-two commensurate unconventional charge density waves~\cite{NayakUDWs,ChakravartyHiddenOrder}. In the latter category, one also finds a charge CI, the so-called chiral $d_{xy}+id_{x^2-y^2}$ density wave state~\cite{Yakovenko90,Tewari,KotetesEPL,KotetesPRBR,ChuanweiZhang,KotetesPRL}. Despite the intense activity, the unambi\-guous verification of a charge CI in cuprates still remains open, thus highlighting the importance of the possible discovery of this exotic phase of matter in Kagome SCs.

\begin{figure*}[t!]
\begin{center}
\includegraphics[width=\textwidth]{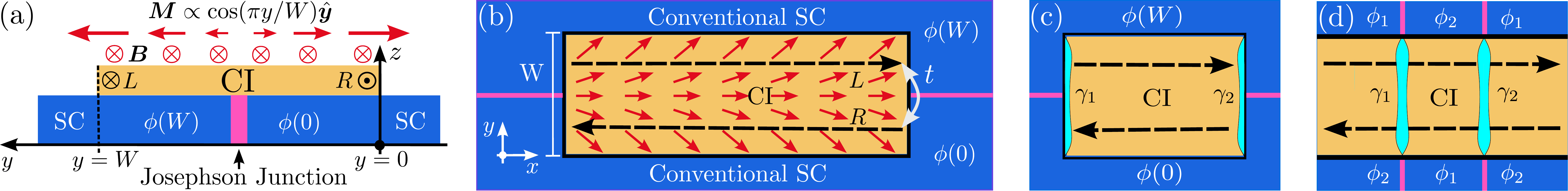}
\end{center}
\caption{(a) Side view of the pyramidal SC/CI/SC hybrid structure with a CI of a width $W$. The two SCs are kept at a phase difference $\delta\phi=\phi(W)-\phi(0)$. This can be experimentally implemented either by means of a Josephson junction or by imposing a supercurrent flow through the two SCs. The CI harbors two counterpropagating chiral edge modes $\{\bm{\otimes},\bm{\odot}\}$. The CI is under the additional influence of a homogeneous magnetic field $\bm{B}=B\hat{\bm{x}}$ and a magnetic stripe $\bm{M}(y)=M\cos[Q(y-y_0)]\hat{\bm{y}}$. The offset $y_0$ controls the strength of the magnetization felt by the two edges. It is crucial that $Q=\big(2\kappa+1\big)\pi/W$ with $\kappa\approx \mathbb{N}^+$ so that $\bm{M}(y)$ is opposite on opposite edges. In our drawing we chose $Q=\pi/W$. Since the properties of the TSC considered here solely stem from the chiral edge modes, it is sufficient for $\bm{B}$ and $\bm{M}(y)$ to be nonzero only near the edges of the CI. (b) Top view of a pyramidal SC/CI/SC hybrid structure with a finite-sized CI strip. The two counterpropagating chiral edge modes (dashed black arrows) become hybridized due to the finite-sized width $W$. The solid red arrows correspond to the magnetization profile which is externally imposed on the CI depicted in (a). (c) The system in (b) harbors two MZMs ($\gamma_{1,2}$) at the two termination edges. The spatial support of the MZMs is sketched with cyan, and is determined by the spatial profile of the edge mode wavefunctions. (d) Alternatively to (c), MZMs can be also trapped in domain walls of the superconducting phase difference $\delta\phi$. We note that for reasons of clarity, the spatial profile of the inhomogeneous magnetic field is not shown in (c) and (d).}
\label{fig:Figure1}
\end{figure*}

At this point we need to emphasize that we employ the name charge CI to also differentiate the state of interest from its close cousin, i.e, the quantum anomalous Hall insulator (QAHI) phase~\cite{QAHIprop1,QAHIprop2}, which also supports a nonzero Chern number and has already been observed in fer\-ro\-ma\-gne\-ti\-cal\-ly doped topological insulators~\cite{QAHIexp}. The experimental discovery of the QAHI also opened perspectives for engineering a topological SC (TSC) through the pro\-xi\-mi\-ty of the QAHI to a conventional SC~\cite{QAHI,LawQuasi1DQAHI}. {\color{black}However, related experiments involving such hybrid sy\-stems did not result in an unambiguous detection of chiral Majorana edge modes~\cite{CMM,CMMM}. In fact, it was shown in Refs.~\onlinecite{WenCMM,SauCMM,NoCMM} that a number of fingerprints which were initially associated with the di\-sco\-ve\-ry of dispersive Majorana modes, could be understood by invo\-king alternative explanations, such as, the good electrical contact between the QAHI and the SC~\cite{WenCMM}, or the pre\-sen\-ce of di\-sor\-der~\cite{SauCMM}. Even more importantly, very recent ex\-pe\-ri\-ments~\cite{Rodenbach,Ferguson,Rosen} have emphasized the urgent need to fully settle whether the bulk or the edge modes are the ones mediating conduction in these systems before they can be employed for functional hybrid devices.}

{\color{black}While the above pending issues may challenge the sui\-ta\-bi\-li\-ty of the QAHIs as platforms to detect the sought-after chiral Majorana edge modes, this general direction of research remains promising also for its potential impact on topological quantum com\-pu\-ting. First of all, it has been theoretically demonstrated~\cite{BiaoLian} that although Majorana chiral edge modes do not adhere to non-Abelian exchange statistics, they can be still harnessed to perform braiding~\cite{KitaevTQC,NayakTQC}. The latter is the quantum operation effected by exchanging two Majorana zero modes (MZMs) in coordinate space~\cite{NayakTQC}. Moreover, it has been also theoretically predicted~\cite{LawQuasi1DQAHI} that coupling two counter\-pro\-pa\-ga\-ting Majorana chiral edge modes engineers systems which can trap MZMs at terminations or domain walls, hence paving the way for versatile Majorana platforms. Therefore, identifying alternative Chern insulators harboring chiral edge modes appears vital for pursuing the above promising research directions.

Given the above hurdles, the possible discovery of CI phases in the AV$_3$Sb$_5$ family of Kagome materials may present a unique opportunity to circumvent drawbacks encountered in QAHIs. Notably, the compatibility of these Kagome materials with both a chiral charge order and superconductivity, may either allow for devices free from the requirement of a proximity effect, or open the door for an improved proximity effect by considering building blocks which originate from the same Kagome material but reside in different phases. Therefore, it is natural to ask what is the me\-cha\-nism that allows conver\-ting a charge CI in coe\-xi\-sten\-ce with conventional superconductivity into a TSC. This appears to be a pres\-sing issue, since the nontrivial topology for both the QAHI and its descendant TSC crucially relies on the pre\-sen\-ce of odd-under-inversion SOC which, however, is not assumed to be present in the charge CI phases of interest. 

In this paper, we answer the above urgent question in the case of the proximity scenario, by uncovering a generic me\-cha\-nism for engineering a TSC in hybrid devices of CIs and conventional SCs. Specifically, we} demonstrate that a TSC becomes accessible by de\-po\-si\-ting a charge CI on top of two conventional SCs kept at a superconducting phase dif\-fe\-ren\-ce. Notably, it is crucial to choose the width of the CI segment in such a manner, so that two counter\-pro\-pa\-ga\-ting electronic chiral edge modes emerge and become weakly hybridized. The CI further needs to be under the in\-fluen\-ce of Rashba-like SOC, which is here assumed to be synthetically ge\-ne\-ra\-ted. In the presence of the net current appea\-ring due to the combination of the edge modes and the super\-con\-ducting phase difference, it is possible to engineer a synthetic Rashba-type SOC by subjecting the CI to an inhomogeneous magnetic field~\cite{KotetesClassi,Heimes,PabloSanJose,Livanas}, which consists of an inplane ferromagnetic component and a transversely-spin-oriented magnetic stripe. The latter is required to feature a suitable periodicity which ensures that the magnetic moment induced by the stripe is pre\-do\-mi\-nan\-tly antiparallel near the edges of the system. {\color{black} Noteworthy, creating such a magnetic configuration appears feasible by means of existing experimental techniques~\cite{Kontos,FrolovMag}.} Blueprints for the hybrid device are presented in Fig.~\ref{fig:Figure1}.

Under the influence of these magnetic fields and the phase-biased superconducting proximity effect, the chiral edge modes transform into a massive Majorana particle~\cite{LawQuasi1DQAHI,MarraMParticle}, which leads to {\color{black}MZMs} at the edges of the system or at mass domain walls. Our analysis identifies the parameter regime in which such a pyramidal SC/CI/SC structure behaves as a MZM track. For this purpose, we adopt an analytical low-energy model based on the chiral edge modes of the CI, and further back our findings by means of exact nu\-me\-ri\-cal stu\-dies on the lattice. Evenmore, we demonstrate that a SC/CI/SC/CI/SC hybrid gives rise to a two-track MZM racetrack which provides a tunable platform for MZM braiding. MZMs can toggle between the two tracks by controlling the intertrack electron tunneling, while they can propagate along each track by spatially varying the superconducting phase differences. Since in these racetracks braiding relies on the MZM track exchange, we put forward spectroscopic methods to experimentally detect the successful implementation of this process. Spe\-ci\-fi\-cal\-ly, by taking into account the particular geometric characteristics of the system, we propose to employ the measurement of the quadrupolar differential conductance which can be defined for a MZM pair. As we show, after time-averaging, this quantity presents certain cha\-ra\-cte\-ri\-stic scaling and quantization features which are in principle experimentally observable. Finally, we identify the experimental knobs that allow fusing two MZMs, while we also bring to the fore an approach for the topological detection of the twofold ground state degeneracy of the system for two uncoupled MZMs.

{\color{black} Before proceeding with our main discussion, we wish to stress that this paper exa\-mi\-nes the experimentally most demanding scenario where Rashba-like SOC is fully absent and the microscopic coexistence of chiral charge order and superconductivity is not feasible. In more convenient situations where the candidate system further exhibits such a SOC or a coexistence, a number of experimental requirements discussed throughout this work are expected to be relaxed. For instance, a magnetic stripe is no longer required when the system is dictated by Rashba SOC, under the condition that this is not key for the topological properties of the CI as it happens for a QAHI. Rashba SOC is typically non-negligible in hybrid devices due to the pre\-sen\-ce of the interface and the concomitant structural inversion asymmetry this incurs. On the other hand, odd-under-inversion SOC is also accessible in bulk systems, either due to the presence of a substrate (Rashba effect) or due to bulk inversion asymmetry (Dresselhaus effect). Evenmore, in the case of a bulk system exhi\-bi\-ting the microscopic coexistence of chiral charge order and superconductivity, the requirement for two phase-biased SCs in the device shown in Fig.~\ref{fig:Figure1} can be correspondingly satisfied by experimentally impo\-sing a gradient on the superconducting phase of the sample.
}

The remainder is organized as follows. In Sec.~\ref{sec:II} we discuss the model Hamiltonian for a single MZM track arising in a SC/CI/SC hybrid structure. Section~\ref{sec:III} contains a low-energy analysis of the model and provides analytical predictions for the topological phase diagram. Section~\ref{sec:IV} continues with the discussion of two coupled MZM tracks in a SC/CI/SC/CI/SC hybrid. As we show in Sec.~\ref{sec:V}, a pair of MZMs can be trapped at a mass domain wall, and can be employed for braiding as detailed in Sec.~\ref{sec:VI}. Sections~\ref{sec:VII} and~\ref{sec:VIII} discuss experimental routes to detect the track exchange and fusion of MZM pairs. We summarize our results and provide an outlook in Sec.~\ref{sec:IX}. Finally, supporting technical details and information, {\color{black}as well as numerical verifications of our ana\-ly\-ti\-cal results, are given in Appendices~A-D.}

\section{Single Track - Model Hamiltonian}\label{sec:II}

In the remainder, we consider a representative model for the charge CI, which in the formalism of second quantization is expressed in terms of the operator $H_{\rm CI}=\int d\bm{k}\ph\bm{\psi}^\dag(\bm{k})\hat{H}_{\rm CI}(\bm{k})\bm{\psi}(\bm{k})$, with the matrix Hamiltonian:
\bea
\hat{H}_{\rm CI}(\bm{k})=\big[\alpha k_x\rho_1+\beta k_y\rho_2+m(k)\rho_3\big]\otimes\mathds{1}_\sigma\,,
\label{eq:Hamiltonian_H_ci}
\eea

\noi where we introduced the wave vector $\bm{k}=(k_x,k_y)$ and its modulus $k=|\bm{k}|$. We consider for convenience that $\alpha,\beta>0$, while we set $m(k)=k^2-m_0$ and $\hbar=1$. The above Hamiltonian is defined in the basis of the spinor:
\bea
\bm{\psi}^\dag(\bm{k})=\big(\psi_{+,\uparrow}^\dag(\bm{k}),\,
\psi_{+,\downarrow}^\dag(\bm{k}),\,\psi_{-,\uparrow}^\dag(\bm{k}),\psi_{-,\downarrow}^\dag(\bm{k})\big)\,.\quad
\eea

\noi In the above, $\pm$ distinguishes between two quantum numbers, such as those of two atomic orbitals of different pa\-ri\-ty, e.g., $s$- and $p$-type. To represent the Hamiltonian, we used the Pauli matrices $\bm{\rho}$ ($\bm{\sigma})$ and the related unit matrix $\mathds{1}_{\rho}$ ($\mathds{1}_{\sigma}$), which act in orbital $\pm$ (spin $\uparrow,\downarrow$) space.

When $m_0>0$, the CI is in the topologically non\-tri\-vial phase. As a consequence, when the CI extends infinitely along the $x/y$ axis it harbors two chiral edge modes which propagate along the $x/y$ axis with one of these re\-si\-ding on the left ($L$) and the other on the right ($R$) edge. With no loss of generality, in the remainder we consider a CI in a strip geometry with a finite width $W$ in the $y$ direction, as shown in Fig.~\ref{fig:Figure1}. Given the chosen geometry, the edge modes feature an energy spectrum which assumes a linear form $E_{L/R}(k_x)=\pm\alpha k_x$ for $k_x\in(-k_c,k_c)$. $k_c$ is a cutoff wave number that controls the validity of the li\-near dispersion. Under the above setup assumption, the chiral edge mode eigenvectors are eigenstates of $\rho_1$. The protected crossing point at $k_x=0$ appears only for $W\rightarrow\infty$. In contrast, for a finite-sized system the chiral modes on opposite edges hybridize. In the remainder, we demonstrate how to harness the chiral edge mode mixing in order to engineer a TSC.

Starting from a topologically nontrivial CI, MZMs become accessible after spin-rotational symmetry is broken, all degeneracies are lifted, and a super\-con\-ducting gap is induced by means of proximity~\cite{SauProxi,Potter}. For further details underlying the above symmetry requirements we urge the reader to consult Refs.~\onlinecite{AZ} and~\onlinecite{SchnyderClassi}. To achieve the first two, we consider the presence of an inhomogeneous magnetic/exchange field. As it was first shown in Ref.~\onlinecite{Heimes}, the presence of a supercurrent in conjunction with canted antiferromagnetism, i.e., coexisting ferromagnetic and transversely oriented antiferromagnetic fields, engineers a synthetic Rashba-like SOC, thus enabling the emergence of MZMs. Motivated by the above, we here assume the presence of an orbital-unselective inhomogeneous field which is described by the Hamiltonian:
\begin{align}
H_{\rm mag}=\int_{-k_c}^{+k_c} dk_x\int_0^W dy\ph\bm{\psi}^\dag(k_x,y)\hat{H}_{\rm mag}(k_x,y)\bm{\psi}(k_x,y)
\end{align}

\noi with the matrix Hamiltonian describing the simultaneous coupling to the magnetization/magnetic fields:
\begin{align}
\hat{H}_{\rm mag}(k_x,y)=\mathds{1}_\rho\otimes\Big\{\bm{M}\cos\big[Q\big(y-y_0\big)\big]-\bm{B}\Big\}\cdot\bm{\sigma}\,.\label{eq:Hmag}
\end{align}

In the above, [$\bm{\psi}(k_x,y)$] $\bm{\psi}(\bm{r})$ is obtained from $\bm{\psi}(\bm{k})$ by means of a [partial] conti\-nuous Fourier transform, with $\bm{r}=(x,y)$ being the real space position vector. The CI is here under the influence of a ferromagnetic component $\propto \bm{B}$, and a magnetic stripe $\propto \bm{M}$ with a mo\-dulation wave number $Q$ and offset $y_0\in\mathbb{R}$. After the mechanism of Ref.~\cite{Heimes}, we deduce that MZMs can become accessible only when $\bm{B}$ and $\bm{M}$ are primarily mutually orthogonal and the value of $QW/\pi$ is very close to an \textit{odd} integer, so that $\cos[Q(y-y_0)]$ changes signs on opposite edges.

Our theoretical proposal becomes complete by further accounting for a proximity induced pairing gap on the CI. For a given $(k_x,y)$ through the most general pairing Hamiltonian is defined as: 
\begin{align}
\sum_{\rho,\rho'=\pm}\psi_{\rho,\uparrow}^{\dag}(k_x,y)\Delta_{\rho\rho'}(k_x,y)\psi_{\rho',\downarrow}^{\dag}(-k_x,y)+{\rm h.c.}.
\label{eq:pairing}
\end{align}

\noi The effective pairing gap $\Delta_{\rho\rho'}(k_x,y)$ is considered here to be of the spin-singlet type and thus satisfies $\Delta_{\rho\rho'}(k_x,y)=\Delta_{\rho'\rho}(-k_x,y)$. Such a pai\-ring gap is assumed here to originate from the pro\-xi\-mi\-ty of the CI to two superconducting segments, as shown in Fig.~\ref{fig:Figure1}.

In the remainder, we consider the simplest scenario where the information regarding the orbital character of the CI ``averages out'', thus resul\-ting in an orbital unselective pairing term. In addition, since we restrict to the vicinity of $k_x\sim0$, we further drop the $k_x$ dependence of $\Delta_{\rho\rho'}(k_x,y)$ and write it as $\Delta_{\rho\rho'}(y)$. Under the above conditions, we find $\Delta_{+-}(y)=0$ and $\Delta_{++}(y)=\Delta_{--}(y)=|\Delta(y)|e^{i\phi(y)}$. $|\Delta(y)|>0$ denotes the modulus, whose exact $y$ dependence is determined by the pro\-per\-ties of the interfaces~\cite{SauProxi,Potter}. In addition, the spatial profile of the super\-con\-duc\-ting phase $\phi(y)$ is assumed to be such, so that the superconducting phase changes sign on opposite edges. As we show below, kee\-ping the two SCs at a phase dif\-fe\-ren\-ce is crucial for crea\-ting and ma\-ni\-pu\-la\-ting MZMs.

\section{Single Track - Low Energy Chiral Edge Mode Hamiltonian}\label{sec:III}

We commence our analysis by expo\-sing the key ingredients required for engineering MZMs. For this purpose, we employ an effective model buil\-ding upon the chiral edge modes of the CI. The effective low-energy Hamiltonian operator which describes such a single pair of spinful chiral edge modes takes the form: 
\bea
{\cal H}_0=\frac{1}{2}\int_{-k_c}^{+k_c}dk_x\bm{{\cal X}}^\dag(k_x)\hat{\cal H}_0(k_x)\bm{{\cal X}}(k_x)\,,
\eea

\noi where we introduced the respective Bogoliubov - de Gennes (BdG) Hamiltonian $\hat{\cal H}_0(k_x)$ which is given by:
\bea
\hat{\cal H}_0(k_x)=\big(\alpha k_x\mathds{1}_\tau\otimes\eta_3+t\tau_3\otimes\eta_1-\mu\tau_3\otimes\mathds{1}_\eta\big)\otimes\mathds{1}_\sigma\,.\quad
\label{eq:effective_model_0}
\eea

\noi The Hamiltonian $\hat{\cal H}_0(k_x)$ acts on the multi-component creation ope\-ra\-tor $\bm{{\cal X}}^\dag(k_x)=\big(\bm{\chi}^\dag(k_x),\,\bm{\chi}^\intercal(-k_x)\big)$ with:
\bea
\bm{\chi}^\dag(k_x)=\big(\psi_{L,\uparrow}^\dag(k_x),\,
\psi_{L,\downarrow}^\dag(k_x),\,\psi_{R,\uparrow}^\dag(k_x),\psi_{R,\downarrow}^\dag(k_x)\big)\,,\phd
\eea

\noi with $^{\intercal}$ denoting matrix transposition. In the above, $t>0$ denotes the ener\-gy scale for the hybridization of the counterpropagating chiral edge modes due to the finite width $W$ of the CI. To represent the Hamiltonian, we made use of the additional Pauli matrices $\bm{\tau}$ and $\bm{\eta}$ along with their related unit matrices $\mathds{1}_{\tau}$ and $\mathds{1}_{\eta}$, which correspondingly act in Nambu (particle-hole) and edge ($L$-$R$) spaces. For convenience, in the remainder we omit writing all unit matrices and the Kronecker pro\-duct symbol ``$\otimes$''. 

Equation~\eqref{eq:effective_model_0} leads to the eight (including spin de\-ge\-ne\-ra\-cy) eigenergies $\pm\sqrt{(\alpha k_x)^2+t^2}\pm\mu$, which resemble the ones obtained for a massive relati\-vi\-stic spin $\nicefrac{1}{2}$ particle/hole in the pre\-sen\-ce of a che\-mi\-cal potential $\mu$. Here, $\mu$ depends on the details of the proximity effect of the CI to the two SCs, since the latter two act as particle reservoirs. The value of $\mu$ is controlled by the band alignment of the materials employed for the hybrid structure, and the electrostatic environment that the CI is exposed to~\cite{Vuik,AntipovPRX,WoodsSP,MikkelsenPRX,Reeg}. Notably, for a charge CI with a sufficiently large dielectric constant, gate electrodes can be employed to experimentally control its electron density.

To obtain the complete low-energy Hamiltonian, we project the magnetic and pairing matrix terms onto the basis of the two counter\-pro\-pa\-ga\-ting chiral edge modes. With no loss of generality we assume the following spin-space profiles for the magnetic field and magnetization: 
\bea 
\bm{B}=(B,0,0)\ph\phd{\rm and}\ph\phd
\bm{M}=(M_{||},M_\perp\cos\omega,M_\perp\sin\omega),\ph\quad
\eea

\noi with $B>0$. $M_{||}>0$ and $M_\perp>0$ denote the mo\-du\-li for the two magnetic stripe components which are parallel and orthogonal to the direction set by the field $\bm{B}$. 

The projection of the magnetic part of the Hamiltonian onto the chiral edge mode sector results into the following low-energy edge-mode BdG Hamiltonian for a single MZM track:
\bea
\hat{\cal H}_{\rm track}(k_x,\omega,\phi)={\cal O}(\omega,\phi)\hat{\cal H}_{\rm track}(k_x){\cal O}^\dag(\omega,\phi)\,.
\label{eq:effective_model_1_phase}
\eea

\noi The above has been parametrized using the ``center-of-mass''  and difference phases:
\bea
\phi=\frac{\phi(W)+\phi(0)}{2}\quad{\rm and}\quad\delta\phi=\phi(W)-\phi(0)\,,
\eea

\noi along with the matrix: 
\bea
{\cal O}(\omega,\phi)={\rm Exp}[i\tau_3(
\phi-\omega\sigma_1)/2]\,,\label{eq:unitaryTrafo}
\eea

\noi which effects the unitary transformation appearing in Eq.~\eqref{eq:effective_model_1_phase}. In addition, we have introduced the $(\omega,\phi)$-independent spectrum-generating matrix Hamiltonian:
\bea
&&\hat{\cal H}_{\rm track}(k_x)=\alpha k_x\eta_3+t\tau_3\eta_1-\mu\tau_3+\tau_3\big(\tilde{M}_{||}\eta_3-B\big)\sigma_1
\no\\
&&+\tilde{M}_\perp\eta_3\sigma_2-\Delta\cos\big(\delta\phi/2\big)\tau_2\sigma_2-\tilde{\Delta}\sin\big(\delta\phi/2\big)\tau_1\eta_3\sigma_2.\qquad
\label{eq:effective_model_1}
\eea

\noi In the above, $\tilde{\bm{M}}$ denotes the effective magnetization felt by the chiral edge modes, and its modulus depends on the value of the stripe offset parameter $y_0$. Moreover, two proximity-induced superconducting gaps $\Delta,\tilde{\Delta}>0$ appear, as a result of the nonzero overlap of the wavefunctions of the two edge modes.

The Hamiltonian in Eq.~\eqref{eq:effective_model_1} belongs to class D~\cite{AZ,SchnyderClassi} with a charge-conjugation symmetry generated by $\Xi=\tau_1{\cal K}$, where ${\cal K}$ denotes complex conjugation. Hence, to infer the topological phase diagram we employ the $\mathbb{Z}_2$ topological invariant first introduced by Kitaev~\cite{KitaevUnpaired}. We define the so-called Majorana number ${\cal M}$ using the Pfaffian of the skew-symmetric matrix: $\hat{{\cal B}}=\tau_1\hat{\cal H}_{\rm track}(k_x=0)$. Specifically, by setting ${\cal M}={\rm sgn}\big[{\rm Pf}(\hat{{\cal B}})\big]$, we find the expression:
\begin{align}
{\cal M}={\rm sgn}\Big[c_+c_-+\big(a-b\big)^2-b^2-d^2\Big]
\label{eq:MajoranaNumber}
\end{align}

\noi where we made use of the shorthand notations:
\bea
a&=&\Delta^2\cos^2(\delta\phi/2),\,\qquad d=2B\tilde{M}_{||},\\
b&=&\tilde{M}^2+B^2-t^2-\mu^2-\tilde{\Delta}^2\sin^2(\delta\phi/2)\,,\\
c_\pm&=&\tilde{M}^2+(B\pm t)^2-\mu^2-\tilde{\Delta}^2\sin^2(\delta\phi/2)\,.
\eea

\begin{figure}[t!]
\begin{centering}
\includegraphics[width=0.9\columnwidth]{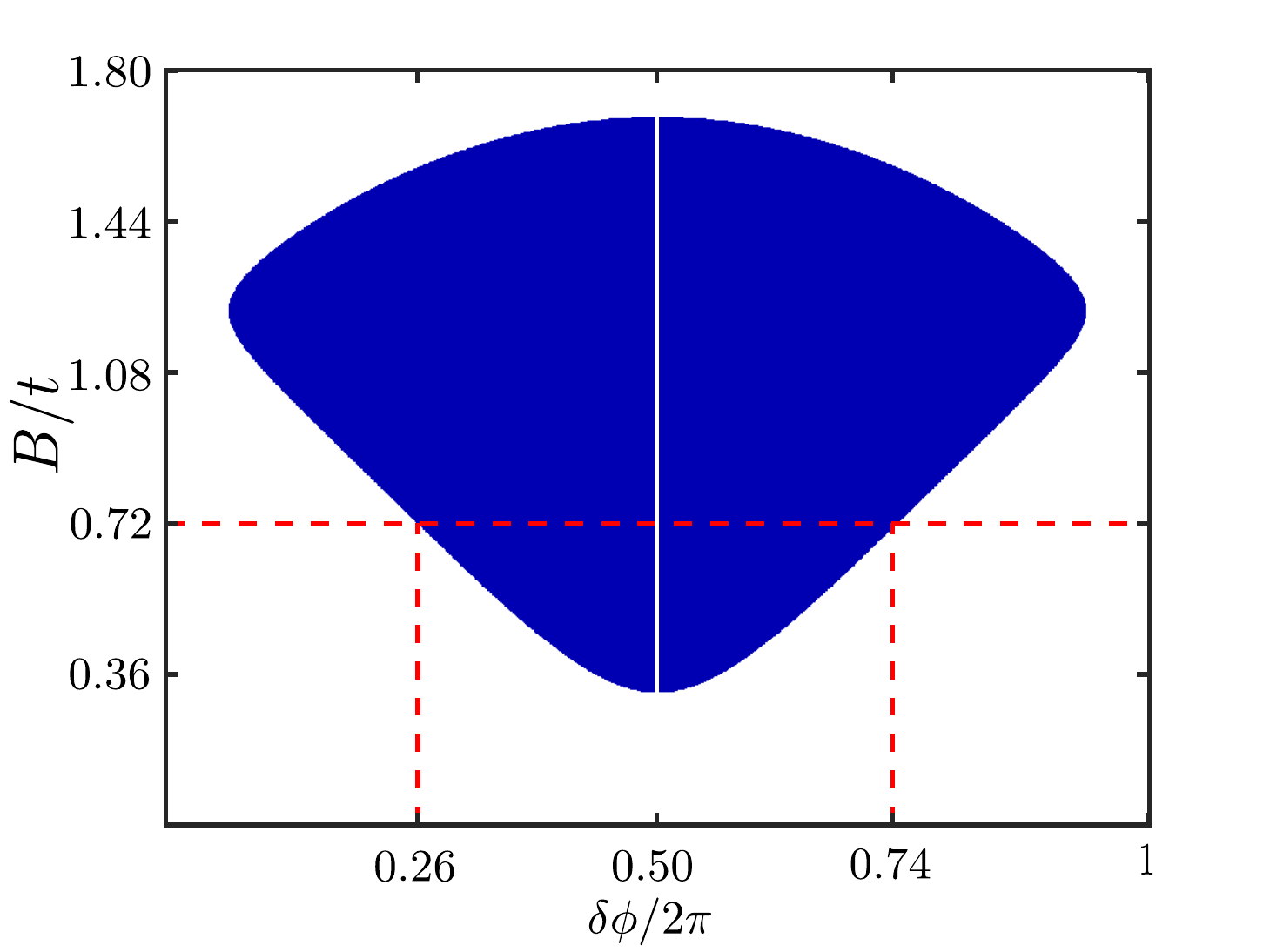}
\par\end{centering}
\caption{Topological phase diagram with respect to the applied magnetic field $B$ and the superconducting phase dif\-fe\-ren\-ce $\delta\phi$. The blue/white region denotes the topologically nontri\-vial/trivial region. The above was extracted using the appro\-xi\-ma\-te analytical Majorana-number expression in Eq.~\eqref{eq:MajoranaNumber}, that we obtained based on the low-energy model of Eq.~\eqref{eq:effective_model_1}. Notably, for $\delta\phi=\pi$ a symmetry class transition occurs which renders the system topologically trivial. The red dashed lines indicate the window for which the system is in the topologically nontrivial phase for $B/t=0.72$, which is a value used for the numerical simulations presented in Fig.~\ref{fig:FigureAppI}. We used the pa\-ra\-me\-ter va\-lues: $\mu=\tilde{M}_{||}=0$, $\Delta/t=0.72$, and $\tilde{M}_{\perp}/t=1.4$.}
\label{fig:Figure2}
\end{figure}

Topological phase transitions occur when ${\cal M}$ changes sign. For a strip of a length $L_x$, MZMs appear at the two terminations edges, as sketched in Fig.~\ref{fig:Figure1}(c). In contrast, as shown in Fig.~\ref{fig:Figure1}(d), in the case of an infinite strip, MZMs can get trapped at domain walls across which ${\cal M}$ changes sign. In Fig.~\ref{fig:Figure2}, we depict the sign-changing be\-ha\-vior of ${\cal M}$ in the $(B,\delta\phi)$ parameter plane, when the re\-mai\-ning parameters are fixed. We observe that both external fields need to reach a thre\-shold value so that a transition to the topologically nontrivial phase emerges.

We remind the reader that, as first discussed by Kitaev~\cite{KitaevUnpaired}, ${\cal M}$ is meaningful as long as the bulk energy spectrum of the system is fully gapped. Hence, in the remainder we restrict to suitably small values of $\tilde{M}_{||}$ which lead to a full gap. Notably, for $\tilde{M}_{||}=0$ the sy\-stem belongs to the BDI symmetry class which allows for multiple MZMs per termination edge protected by a chiral symmetry. Nevertheless, as we prove in App.~\ref{app:AppendixI}, even in the BDI case gap closings can only occur at $k_x=0$. Hence, ${\cal M}$ is sufficient for inferring the topological phase dia\-gram also in this case. We also note that the spectrum of Eq.~\eqref{eq:effective_model_1} appears to be gapless for $\delta\phi=\pi$ for arbitrary values of $\tilde{M}_{||}$. Even more, when we simultaneously consider $\delta\phi=\pi$ and $\mu=0$, the emergence of an extra time-reversal symmetry mediates the symmetry class transition BDI$\rightarrow$AI$\oplus$AI. Notably, the latter class is tri\-vial in 1D~\cite{SchnyderClassi}, and thus prohibits the appearance of MZMs, as we further explain in App.~\ref{app:AppendixI}. Indeed, the above is also corroborated by the results shown in Fig.~\ref{fig:Figure2}. {\color{black}To substantiate the emergence of MZMs more trans\-pa\-rently, we provide in App.~\ref{app:AppendixII} complementary numerical verifications of the results in Fig.~\ref{fig:Figure2}.} 

To fa\-ci\-li\-ta\-te the upcoming discussion of brai\-ding, we here demonstrate that an effective spinless p-wave SC model becomes engineered in a single track. Spe\-ci\-fi\-cal\-ly, we find that when $t$ is the largest ener\-gy scale, $B$ the se\-cond largest, and $\mu>0$, we can project the Hamiltonian in Eq.~\eqref{eq:effective_model_1} onto the $\eta_1=\sigma_1=1$ eigenstate, which is responsible for the nontrivial topology in the given regime of pa\-ra\-me\-ter values. As we detail in App.~\ref{app:AppendixIII}, this projection yields the fol\-lowing spinless p-wave SC model for a single infinitely long track:
\begin{align}
\hat{{\cal H}}_{\rm track}(k_x)=Jk_x+\upsilon k_xe^{i(\phi-\omega)\tau_3}\tau_2+m\tau_3,
\label{eq:SingleTrack}
\end{align}

\noi where we set $\tilde{M}=|\tilde{\bm{M}}|$ and defined the coefficients:
\begin{align}
J=\frac{\alpha\tilde{M}_{||}}{t},\,\quad\upsilon=\frac{\alpha\tilde{M}_\perp\Delta\cos(\delta\phi/2)}{tB},\,\qquad\qquad\no\\
m=t-B-\mu+\frac{t\Delta^2+B\tilde{\Delta}^2}{4Bt}\cos\delta\phi+\frac{2\tilde{M}^2-\tilde{\Delta}^2}{4t}+\frac{\Delta^2}{4B}.\no
\end{align}

Notably, the vanishing of $\upsilon$ for $\delta\phi=\pi$ provides an alternative route to transparently understand the trivial character of the system for this phase dif\-fe\-ren\-ce value that was emphasized earlier. Further, we find that the parallel component $\tilde{M}_{||}$ of $\tilde{\bm{M}}$ leads to the induction of net momentum $\propto J$ along the SC/CI/SC strip~\cite{OjanenME}. Finally, in accordance with the Hamiltonian in Eq.~\eqref{eq:effective_model_1_phase}, the orien\-ta\-tion of the orthogonal component defined by $\omega$ effects a unitary transformation, and after the projection solely modifies the ``center-of-mass'' phase $\phi$.

\section{Racetrack - Model Hamiltonian}\label{sec:IV}

\begin{figure}[t!]
\begin{center}
\includegraphics[width=\columnwidth]{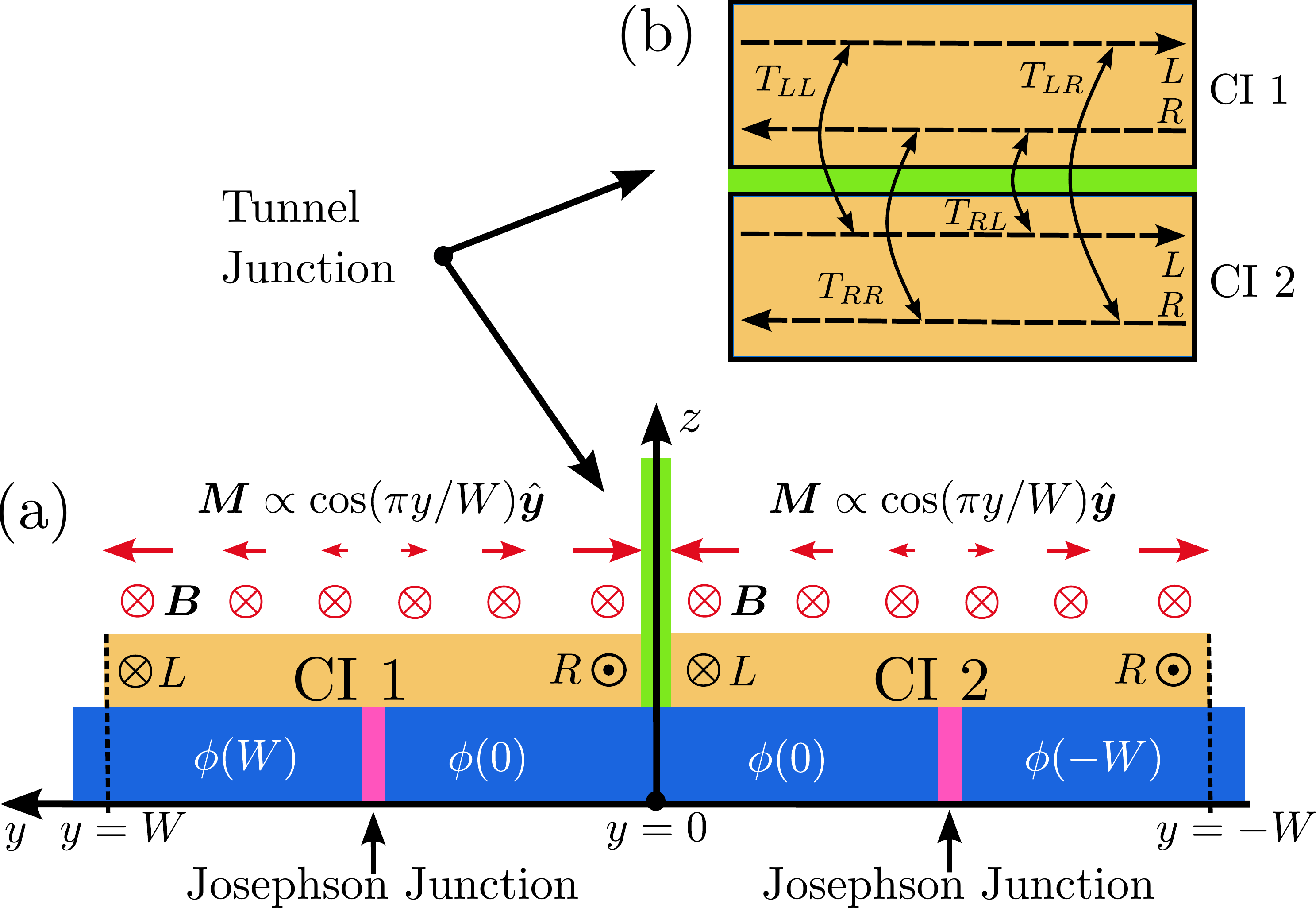}
\end{center}
\caption{(a) Side view of a double MZM track (MZM racetrack) constructed by a double-pyramidal SC/CI/SC/CI/SC hybrid structure. In the most symmetric situation the two CIs are considered to be identical. In order to engineer two effective p-wave SCs, the two phase differences defined for the three SCs, i.e., $\delta\phi_1=\phi(W)-\phi(0)$ and $\delta\phi_2=\phi(0)-\phi(-W)$,  must take suitable values so that two effective p-wave SCs are simultaneously induced. By controlling the electron tunnel coupling among the four chiral edge modes of the two CIs, one can couple the two effective p-wave SCs in a tunable fa\-shion. (b) Top view of the MZM racetrack with a focus on the tunnel junction bringing the two CIs in electronic contact. The various double arrows depict the tunneling processes appearing between pairs of chiral edge modes of the two CIs.}
\label{fig:Figure3}
\end{figure}

Coupling two hybrid structures of the type proposed above opens perspectives for braiding MZMs using racetracks. The minimal setup to implement a racetrack requires three SCs and is sketched in Fig.~\ref{fig:Figure3}. The middle SC segment couples to two identical CIs, while each one of the two CIs are deposited on top of three conventional SCs. Thus, the resulting SC/CI/SC/CI/SC racetrack features the two independent phase dif\-fe\-ren\-ces:
\bea
\delta\phi_1=\phi(W)-\phi(0)\phd {\rm and}\phd \delta\phi_2=\phi(0)-\phi(-W)\,,\phd\phd
\eea

\noi which are defined for the CI denoted with $1,2$. Notably, the arising reference phase $\phi(0)$ of the middle SC can be for convenience set to zero for racetracks in which there exists only a single domain wall harboring a MZM pair.

The racetrack Hamiltonian is obtained by coupling the two effective p-wave SCs which arise from each track. We restrict to couplings originating from low-energy inter\-track tunneling processes between the two pairs of chiral edge modes, which are expressed through the term: 
\bea
{\cal H}_{\rm intertrack}=\int_{-k_c}^{+k_c}dk_x\Big[\bm{\chi}_1^\dag(k_x)\hat{\cal H}_{\rm T}\bm{\chi}_2(k_x)+{\rm h.c.}\Big]\,,\quad
\eea

\noi where we introduced:
\bea
\hat{\cal H}_{\rm T}&=&T_{LL}e^{i\varphi_{LL}}\frac{1+\eta_3}{2}+T_{RR}e^{i\varphi_{RR}}\frac{1-\eta_3}{2}
\no\\
&+&T_{LR}e^{i\varphi_{LR}}\frac{\eta_1+i\eta_2}{2}+T_{RL}e^{i\varphi_{RL}}\frac{\eta_1-i\eta_2}{2}\,.
\label{eq:InterTrackCoupling}
\eea

\noi In the above, the moduli $T_{LL,RR,LR,RL}>0$ control the strengths of the various intertrack tunneling processes, while the phases $\varphi_{LL,RR,LR,RL}$ are only nonzero when flux $\Phi_z$ pierces the cross-section of the two CIs. Since we restrict to the low-energy regime, the va\-rious tunnel couplings have been assumed to be independent of the wave number $k_x$. In this work, we predominantly consider the case of a zero out-of-plane flux, and set the above phases to zero throughout. The only exception is Sec.~\ref{sec:VIII} where we briefly discuss the implications of a nonzero flux in connection to the fusion of MZM pairs. Even more, in the remainder we also assume that the tunnel matrix elements between modes of the same type, i.e., $LL$ and $RR$, are equal and hence set $T_{RR}=T_{LL}=T_S$ from now on.

We now proceed by extending the spinor $\bm{{\cal X}}^\dag(k_x)$ to the two-component superspinor $\big(\bm{{\cal X}}_1^\dag(k_x),\bm{{\cal X}}_2^\dag(k_x)\big)$ in order to take into account the upper and lower track degrees of freedom. Putting together the inter\-track tunnel coupling Hamiltonian of Eq.~\eqref{eq:InterTrackCoupling} and the two single-track Hamiltonians which are expressed following Eq.~\eqref{eq:effective_model_1_phase}, yields the BdG Hamiltonian below:
\begin{align}
\hat{\cal H}_{\rm racetrack}(k_x,\omega_{1,2},\phi_{1,2})=
{\cal O}(\Omega,\Phi)
\hat{\cal H}_{\rm racetrack}(k_x){\cal O}^\dag(\Omega,\Phi)\label{eq:RacetrackParamParam}
\end{align}

\noi where we introduced the racetrack ``center-of-mass'' va\-ria\-bles: 
\bea
\Omega&=&\frac{\omega_1+\omega_2}{2}\,,\\
\Phi&=&\frac{\phi_1+\phi_2}{2}=\frac{\phi(0)}{2}+\frac{\phi(W)+\phi(-W)}{4}
\eea

\noi as well as the associated matrix Hamiltonian in the extended Hilbert space:
\bea
&&\hat{\cal H}_{\rm racetrack}(k_x)=\sum_{\lambda=\pm}
\hat{\cal P}_\lambda\hat{{\cal H}}_{\rm track}^\lambda(k_x)\no\\
&&+{\rm Exp}\left\{i\lambda_3\tau_3\left[\frac{\omega_1-\omega_2}{2}\sigma_1-\frac{\phi(W)-\phi(-W)}{4}\right]\right\}\bigg[T_S\lambda_1\tau_3\no\\
&&+\big(T_{RL}+T_{LR}\big)\lambda_1\tau_3\eta_1/2+\big(T_{RL}-T_{LR}\big)\lambda_2\tau_3\eta_2/2\bigg]
\label{eq:RaceTrack}
\eea

\noi where we made use of the projectors $\hat{\cal P}_\pm=(1\pm\lambda_3)/2$ onto the track labelled by 1 and 2. In addition, we made a convenient gauge choice, so that the arising difference in the phase factors ${\rm Exp}[i\tau_3(
\phi_{1,2}-\omega_{1,2}\sigma_1)/2]$ of the Hamiltonians for each track, enters in the inter\-track couplings. 

We remark that the phases $\Omega$ and $\Phi$ do not influence the energy spectrum and, thus, can be set to zero for our upcoming analysis. However, these center-of-mass phases become important when considering double domain walls harboring four MZMs in total~\cite{Sticlet2013,PKsynthetic,MTMPRB}. In such situations $\Omega$ and $\Phi$ can vary in space and, thus, give rise to Josephson junctions which enable the observation of a number of unusual current responses stemming from the two underlying pairs of MZMs~\cite{Sticlet2013,PKsynthetic,MTMPRB}. These include the emergence of chiral anomaly and the emergence of Weyl points in a synthetic space~\cite{PKsynthetic,MTMPRB}. In fact, the search for Weyl points has recently attracted significant attention in the context of multi-terminal conventional~\cite{Riwar,Eriksson,Meyer_PRL,LevchenkoI,LevchenkoII,Belzig,Rastelli,WeylCircuits,LevchenkoNonAbelian} and topological~\cite{PKsynthetic,MTMPRB,Mi,Balseiro,Sakurai,Houzet} Josephson junctions~\cite{Draelos, Manucharyan,Arnault}.    

Along the lines of the process that led to Eq.~\eqref{eq:SingleTrack}, we proceed by here projecting the racetrack Hamiltonian in Eq.~\eqref{eq:RaceTrack} onto the $\eta_1=\sigma_1=1$ eigenstate, we obtain a low-energy model which describes two coupled single-track effective spinless p-wave SCs. While a detailed analysis of this procedure is presented in App.~\ref{app:AppendixIII}, in the present section we restrict to the most symmetric scenario, in which the two coupled p-wave SCs feature different mass terms $m_{1,2}$ but are other\-wi\-se identical. Under such a condition, the BdG Hamiltonian for an infinitely long MZM racetrack becomes:
\bea
\hat{{\cal H}}_{\rm racetrack}(k_x)=\upsilon k_x\tau_2+\big(m_++m_-\lambda_3+T\lambda_1\big)\tau_3,\phd\quad
\label{eq:RaceTrackSym}
\eea

\noi where we employed the compact notation:
\bea
m_\pm&=&\frac{m_1\pm m_2}{2}\quad{\rm and}\quad T=T_S+\frac{T_{LR}+T_{RL}}{2}\,.\phd\quad
\eea

\noi Moreover, we considered for simplicity that $M_{||,1}=M_{||,2}=0$, which in turn resulted in $J_1=J_2=0$.

\section{MZM Pair at a Mass Domain Wall}\label{sec:V}

\begin{figure}[t!]
\begin{centering}
\includegraphics[width=\columnwidth]{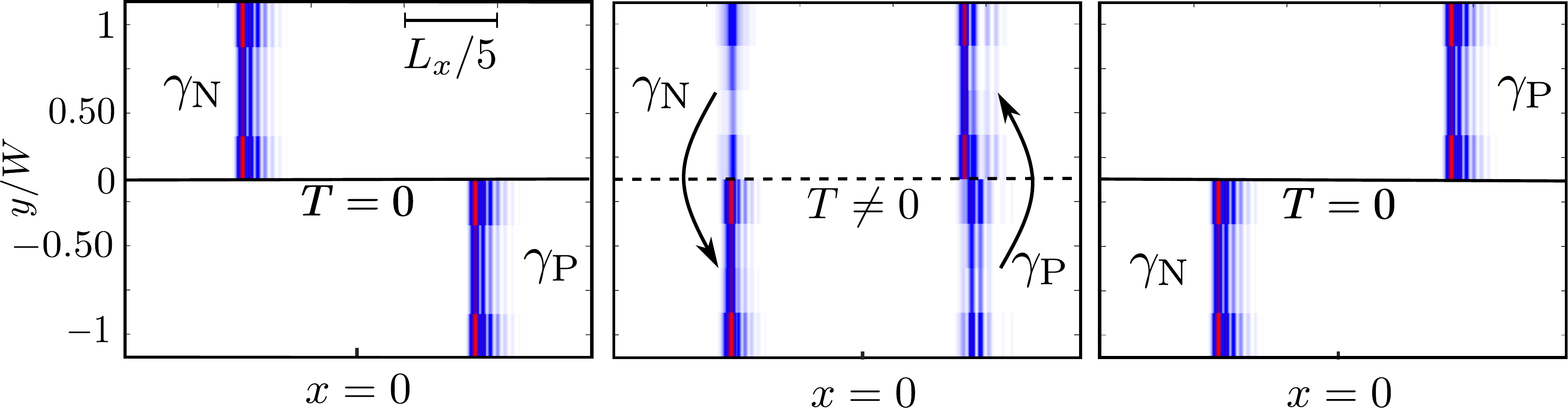}
\par\end{centering}
\caption{Numerical simulation of instants of the MZM track exchange process. Initially, see (a), the two tracks are decoupled and each one harbors a single MZM ($\gamma_{\rm N,P}$) at a domain wall controlled by the arising two superconducting phase differences $\delta\phi_{1,2}$. (b) is obtained by switching on electron tunneling across the two tracks ($T\neq0$). Each MZM has now spatial support on both tracks. The superconducting phases and tunnel coupling strength need to be continuously varied to carry out the complete track exchange depicted in (c). For the above we employed the parameters: $L_x=2400$, $W=8$, $|M_{\perp}|=0.02$, $|M_{||}|=\mu=0$, $m_0=1$, $\alpha=\beta=1$. For $\bm{M}(R_y)$ and $\phi(R_y)$, we adopted the spatial profiles defined in App.~\ref{app:AppendixII}.}
\label{fig:Figure4}
\end{figure}

Given the above low-energy Hamiltonian, we proceed by considering the presence of a domain wall which stabilizes a MZM pair per racetrack, cf Fig.~\ref{fig:Figure4}, whose braiding we exa\-mi\-ne later on. To engineer a pair of MZMs, the masses $m_\pm$ are required to be spatially varying. For convenience and with no loss of generality, we consider a domain wall centered at $x=0$, which has the following  spatial profile:
\bea
m_1(x)=m_0\frac{x+x_0}{\xi_{\rm dw}}\quad{\rm and}\quad 
m_2(x)=m_0\frac{x-x_0}{\xi_{\rm dw}}\,,\quad
\eea

\noi where $\xi_{\rm dw}$ defines the spatial extent of the domain wall. The above expressions imply that when the two tracks are decoupled, i.e. for $T=0$, track 1 (2) harbors a MZM at position $x=-x_0$ ($x=x_0$). This is straightforward to obtain by accordingly extending the model Hamiltonian of Eq.~\eqref{eq:RaceTrackSym} to its coordinate space counterpart given by:
\bea
\hat{{\cal H}}_{\rm racetrack}^{\rm dw}(\hat{p}_x,x)=\upsilon\hat{p}_x\tau_2+m_0\frac{x+x_0\lambda_3+x_T\lambda_1}{\xi_{\rm dw}}\tau_3,\phd\quad
\label{eq:RaceTrackDW}
\eea

\noi where we defined the lengthscale $x_T=T\xi_{\rm dw}/m_0$ which quantifies the strength of the inter\-track tunnel coupling. 

In the above, the term $x_0\lambda_3+x_T\lambda_1$ can be readily dia\-go\-na\-li\-zed by introducing its eigenstates in track $\lambda$ space:
\bea
\left|u_{\rm N}(\theta)\right>=\left(\begin{array}{c}\cos(\theta/2)\\\sin(\theta/2)\end{array}\right),\,\,
\left|u_{\rm P}(\theta)\right>=\left(\begin{array}{c}\phd\ph\sin(\theta/2)\\-\cos(\theta/2)\end{array}\right)\phd\,\quad
\label{eq:RacetrackEigen}
\eea

\noi with respective eigenvalues $\pm x_{\rm MZM}$, where $x_{\rm MZM}=\sqrt{x_0^2+x_T^2}$. The angle angle $\theta$ is defined through the relation $\tan\theta=x_T/x_0$. We note that the solution $\left|u_{\rm N}(\theta)\right>$ [$\left|u_{\rm P}(\theta)\right>$] always describes the MZM which is located on the negative [po\-si\-ti\-ve] side of the $x$ axis at position $\mp x_{\rm MZM}$. For $\theta=0$ the MZM on the negative (positive) side appears on track 1 (2), as depicted in Fig.~\ref{fig:Figure4}(a). Instead, for $\theta=\pi$ the MZM exchange tracks compared to $\theta=0$. See as depicted in Fig.~\ref{fig:Figure4}(c) for a sketch. Re\-markably, processes that adiabatically modify $\theta$ in the interval $\theta\in[0,\pi]$ perform a track exchange for the MZM on a given ne\-ga\-tive/positive side on the $x$ axis. As we discuss in the next paragraphs, this exchange presented in Fig.~\ref{fig:Figure4}(b), is crucial for the braiding of a pair of MZM in a two-track racetrack. 

We proceed with obtaining the eigenvectors of the MZMs. After introducing the eigenstates in Eq.~\eqref{eq:RacetrackEigen}, the Hamiltonian for the MZM at $\mp x_{\rm MZM}$ is proportional to the operator: $\xi_{\rm dw}\upsilon\hat{p}_x\tau_2+m_0(x\pm x_{\rm MZM})\tau_3$. Therefore, the pair of MZMs appears as the zeroth Landau level solution of each massless Dirac Hamiltonian in the presence of an effective magnetic field~\cite{GrapheneRMP}. Each MZM eigenvector is required to be normalizable, which is the condition that enforces that the MZM eigenvectors are eigenstates of the chiral symmetry operator $\tau_1$ of a specific chirality $\tau_1=\pm1$. For $\upsilon,m_0>0$ we find that both MZMs have chirality $\tau_1=-1$ for the given properties of the domain wall. Finally, after accounting for the spatial parts of the MZM eigenvectors we find that these assume the form:
\bea
\left|\gamma_{\rm N/P}(x,\theta)\right>=f_{\rm N/P}(x)\left|u_{\rm N/P}(\theta)\right>\otimes\left|\tau_1=-1\right>,\,\,\quad
\label{eq:MZMeigenvectors}
\eea

\noi with the respective spatial wavefunction distribution:
\bea
f_{\rm N/P}(x)\propto{\rm Exp}\left[-\frac{(x\pm x_{\rm MZM})^2}{2\upsilon\hbar\xi_{\rm dw}/m_0}\right]\,.
\eea

\noi Here, each one of the spatial profiles shown above needs to be normalized in such a way so that the following defining relations for the Majorana operators hold: 
\bea
\{\gamma_{\rm N},\gamma_{\rm N}\}=\{\gamma_{\rm P},\gamma_{\rm P}\}=1\phd{\rm  and}\phd \{\gamma_{\rm N},\gamma_{\rm P}\}=0.\phd
\eea

We conclude the section with commenting on the protection of the MZM pair trapped at such a domain wall. First of all, the emergence of the two MZMs is straightforward to understand when the two racetracks are completely decoupled. As long as inter-track tunneling is prohibited, the two MZMs remain uncoupled even if they have spatial support at the same region. However, switching on the  inter-track tunneling in a region where the wavefunctions of both MZMs have a nonzero spatial support, generally degrades the robustness of the MZM pair. The fate of the MZM pair depends on whether the Hamiltonian in Eq.~\eqref{eq:RaceTrack} possesses a chiral symmetry or not, since the former scenario allows for multiple uncoupled MZMs per domain wall. When the tunnel matrix elements are real, the MZM pair is preserved due to the emergence of a chiral symmetry with matrix $\Pi=\tau_1$. However, terms which violate this symmetry hybridize the MZM pair into nonzero energy Andreev modes. This is in fact what happens when a nonzero out-of-plane flux $\Phi_z$ threads the tunnel junction and renders the inter-track tunnel matrix elements complex. In this case a term $\propto\lambda_2$ is added to Eq.~\eqref{eq:RaceTrackDW} which violates $\Pi$ and mixes the MZMs of the pair. In Sec.~\ref{sec:VIII} we demonstrate how one can actually exploit this property for fusing MZMs and inferring information regarding the two-fold ground state degeneracy of a domain wall which harbors a MZM pair.

\section{MZM Braiding in a Racetrack}\label{sec:VI}

\begin{figure}[t!]
\begin{center}
\includegraphics[width=\columnwidth]{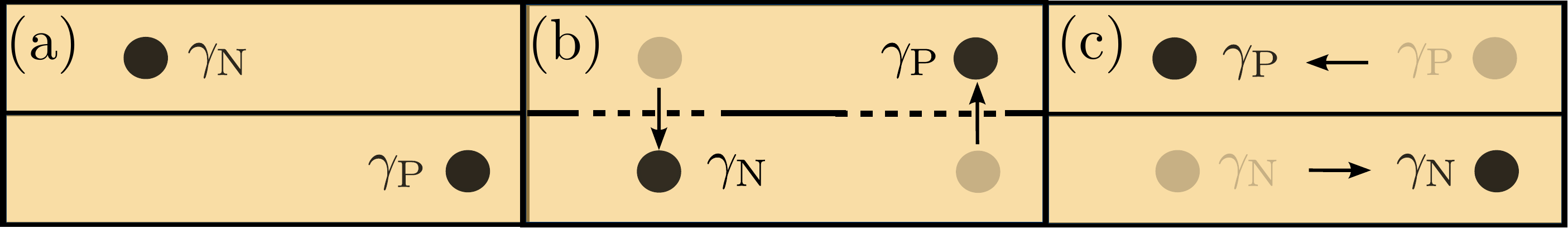}
\end{center}
\caption{MZM braiding in a racetrack. (a) Initially, two MZMs $\gamma_{\rm N,P}$ are trapped by a mass domain wall, as described in Eq.~\eqref{eq:RaceTrackDW}. (b) Afterwards, one adiabatically switches on and further varies the transparency of the junction which mediates intertrack electron tunneling. Simultaneously, the superconducting phase differences are spatially adjusted so that MZM track exchange is implemented. (c) After the track exchange, intertrack tunneling is switched off, and the two MZMs get shuttled by spatially varying the phases differences in an adia\-ba\-ti\-c manner, in order for $\gamma_{\rm N,P}$ to be exchanged.}
\label{fig:Figure5}
\end{figure}

In this section, we proceed with putting forward a protocol that allows for the spatial exchange of two MZMs in the racetrack. The basic principle is illustrated in Fig.~\ref{fig:Figure5}. 

At first, a single MZM is created in each track as shown in Fig.~\ref{fig:Figure5}(a). In the vicinity of the MZMs, we impose $T=0$ which implies $\theta=0$. At the same time, the  intertrack coupling is required to be nonzero far away from the two MZMs, in order to guarantee that the two tracks remain electronically connected during the braiding process. This is a crucial requirement for keeping a common gauge for the two tracks, eventhough the two MZMs may belong two different tracks. 

Next, one ramps up $T$ at the positions $\pm x_{\rm MZM}$ so to allow the MZMs to tunnel through the interface of the two tracks and begin the exchange process. See panels (b) of Figs.~\ref{fig:Figure4} and~~\ref{fig:Figure5}. In order to describe the track exchange process it is instructive to monitor the evolution of the MZM eigenvectors in Eq.~\eqref{eq:MZMeigenvectors}. In the minimal protocol one needs to suitably adjust $x_0$ and $x_T$ during the track exchange, so that $x_{\rm MZM}$ remains unaltered. Equi\-va\-lently, this implies that one is required to adia\-ba\-ti\-cal\-ly vary $\theta$ in the interval $[0,\pi]$. Indeed, as we verify from our nu\-me\-ri\-cal results in Fig.~\ref{fig:Figure4}(b) for a value of $\theta$ in this interval, the MZM are now in a superposition state with weight in both tracks. When $\theta$ becomes $\pi$, each MZM finds itself in a track different than the one that it was located for $\theta=0$. Remarkably, an adiabatic process that leads to $\theta\mapsto\theta+\pi$ implies the following transformation property for the MZM eigenvectors:
\begin{align}
\left|u_{\rm N}(\theta+\pi)\right>=-\left|u_{\rm P}(\theta)\right>
\phd{\rm and}\phd
\left|u_{\rm P}(\theta+\pi)\right>=+\left|u_{\rm N}(\theta)\right>.
\label{eq:MZM_Eigenvector_Trafo}
\end{align}

\noi Interestingly, the above transformation behaviour is identical to the one that we expect to obtain for the MZM operators $\gamma_{\rm N/P}$ at the end of the braiding process, i.e., $\gamma_{\rm N}\mapsto-\gamma_{\rm P}$ and $\gamma_{\rm P}\mapsto+\gamma_{\rm N}$. In fact, as we argue below and further prove in App.~\ref{app:AppendixIV}, given the chosen gauge, it is exactly the track-exchange part that leads to the desired braiding transformation properties for the MZMs.

The third and final step that is needed to complete the braiding operation is to exchange the two MZM positions, as depicted in Fig.~\ref{fig:Figure5}(c). Keeping $\theta$ fixed at $\pi$, which essentially implies that the two tracks should be decoupled in the region where the MZMs are located, one needs to adiabatically change the sign of $x_{\rm MZM}$. Hence, at the end of this position exchange, the spatial part of the MZM eigenvectors get swapped. As we show in App.~\ref{app:AppendixIV}, this final stage of the braiding process does not introduce any further re\-la\-ti\-ve sign changes for the MZM operators. Within the framework chosen here, the sign changes that appear du\-ring brai\-ding so\-le\-ly stem from the track exchange. In part, this should be anticipated since the Hamiltonian is real and the appearance of Berry phases should be attributed to the mismatch arising in the MZM eigenvectors at the beginning and end of the process~\cite{AliceaTQC}. This is indeed reflected in the transformation result of Eq.~\eqref{eq:MZM_Eigenvector_Trafo}. Nonetheless, to concretely support this argument, we prove the above in App.~\ref{app:AppendixIV} using a topologically-equivalent two-track spinless Kitaev chain model~\cite{KitaevUnpaired}.

\section{MZM Track Exchange - Experimental Signatures}\label{sec:VII}

As mentioned in the previous paragraph, the most crucial part of braiding is the track exchange that the MZMs undergo. Hence, in order to gain a high-level of control over the MZM braiding, it appears imperative to be in a position to experimentally verify the proper completion of the MZM track exchange process. Since during braiding the two MZMs are considered to be separated infinitely apart, one can reside on the commonly used MZM spectroscopic probes. However, as we bring forward here, the present MZM platform opens perspectives for new types of spectroscopic measurement approaches.

\begin{figure}[t!]
\begin{center}
\includegraphics[width=0.48\columnwidth]{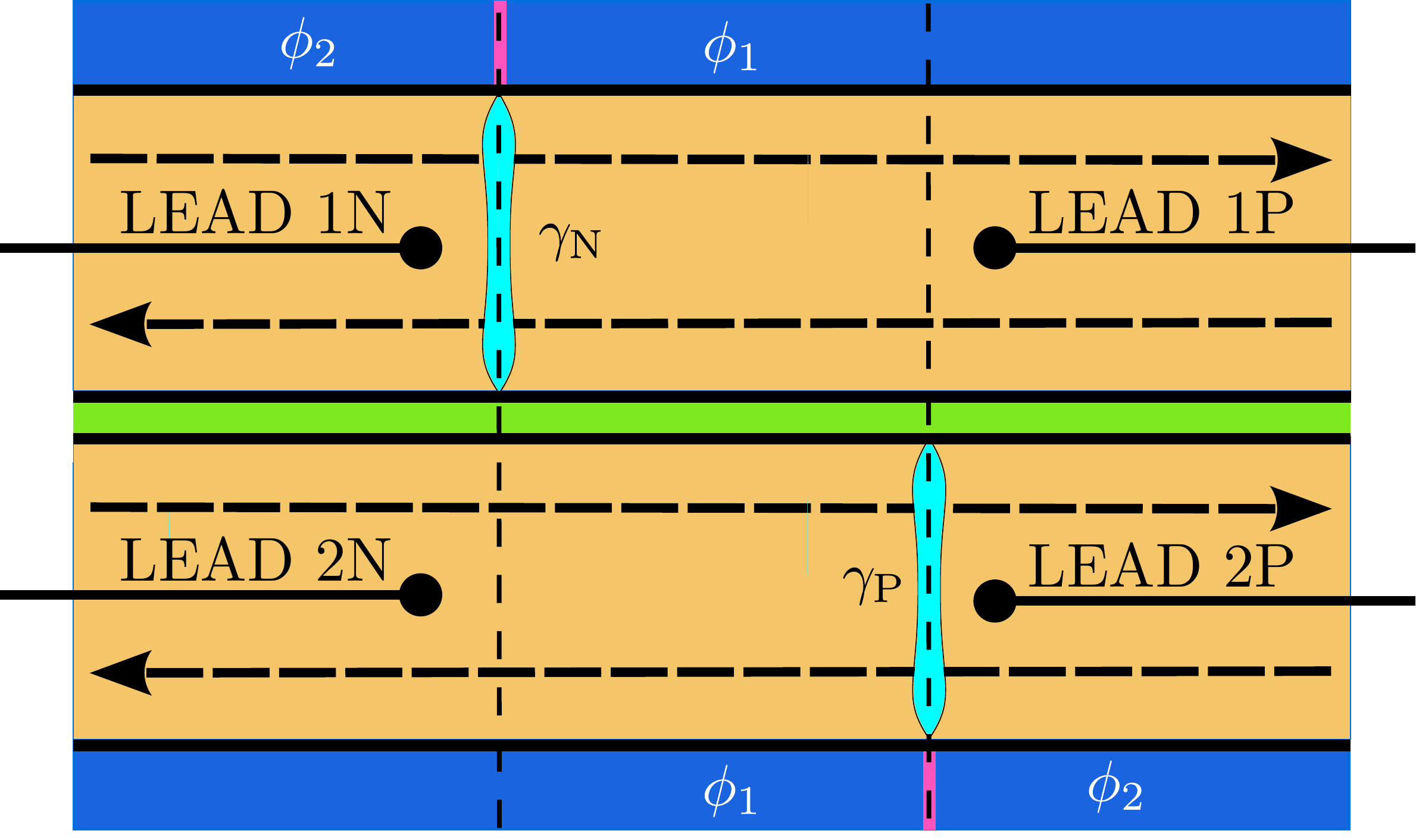}\hspace{0.03in}
\includegraphics[width=0.50\columnwidth]{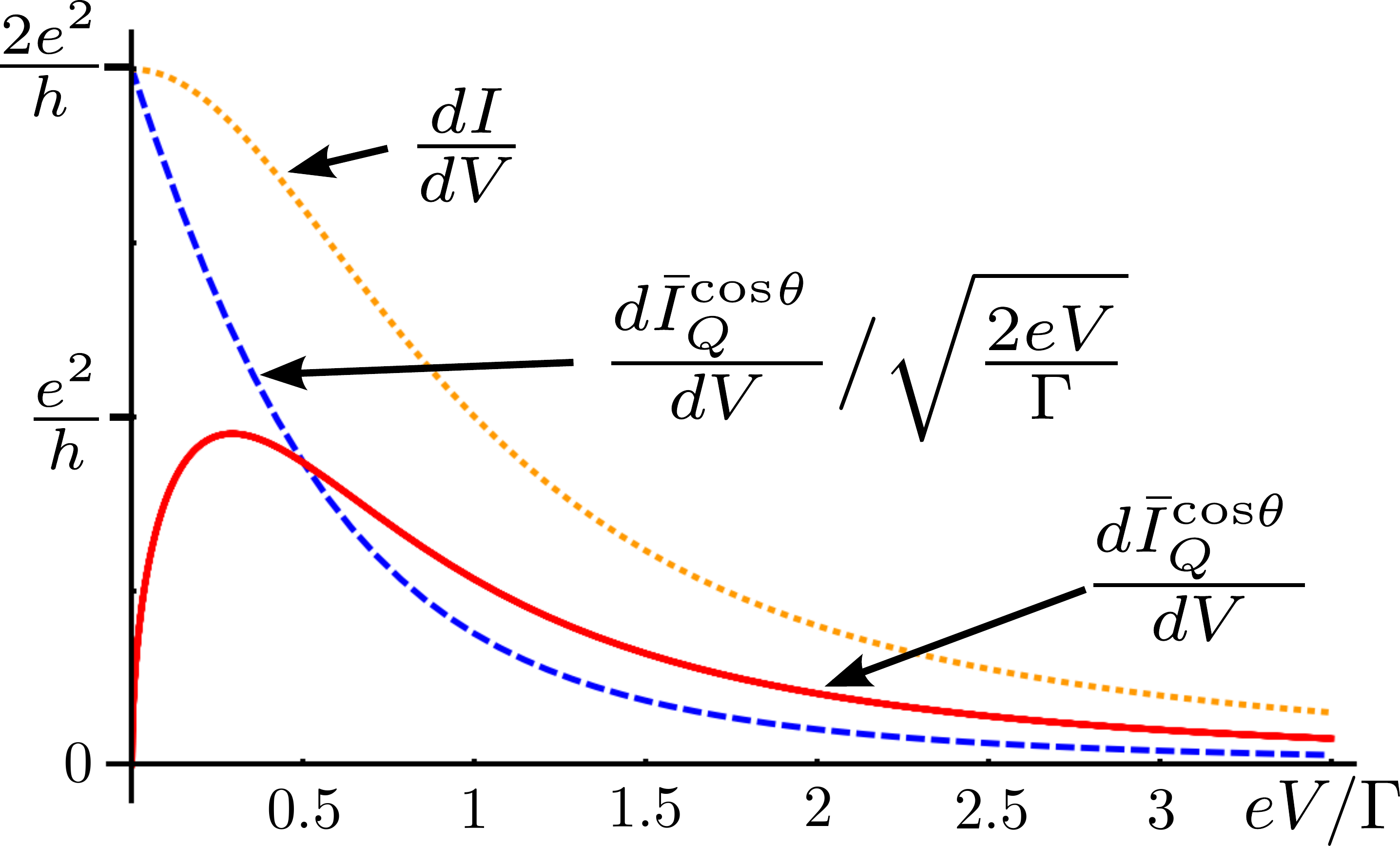}
\end{center}
\caption{(a) Experimental setup to verify the MZM track exchange. Four normal leads are attached to the two CIs of the racetrack. The leads' tips are placed near the four locations where the two MZMs are expected to appear during the MZM track exchange. We depict the initial positions of the two MZMs. After the MZM track exchange is completed, the two MZMs are expected to appear at the positions of leads 1P and 2N. (b) Single-lead differential conductance $dI/dV$, and the $\cos\theta$ component $d\bar{I}_Q^{\cos\theta}/dV$ of the time-averaged quadrupolar dif\-fe\-ren\-tial conductance. After di\-vi\-ding by $\sqrt{2eV/\Gamma}$, the latter becomes equal to $2e^2/h$ at $V=0$, similar to $dI/dV$.}
\label{fig:Figure6}
\end{figure}

As depicted in Fig.~\ref{fig:Figure6}(a), we consider that the platform is attached to four metallic leads. These contact the sy\-stem exactly at the four points of coordinate space where the two MZMs appear at the beginning and at the end of the track exchange. As it is well established, the dif\-fe\-ren\-tial conductance $dI/dV$ measured at zero bias voltage ($V=0$) by a single lead coupled to a single MZM, exhibits a characteristic peak~\cite{SauZBP} which is quantized and equal to $2e^2/h$~\cite{LawZBP,FlensbergZBP}. Here, by restricting to the MZM pair subspace, we find that we can define four distinct differential conductance values which, when the leads are identical and kept at the same potential, they become spatially inter-related in a specific fashion. Motivated by this observation, we introduce the quadrupolar differential conductance for the domain wall region, defined as:
\begin{align}
\frac{dI_Q}{dV}=\frac{1}{4}\left(\frac{dI_{1,{\rm N}}}{dV}-\frac{dI_{2,{\rm N}}}{dV}-\frac{dI_{1,{\rm P}}}{dV}+\frac{dI_{2,{\rm P}}}{dV}\right),
\end{align}

\noi where the indices ${\rm N/P}$ and $1/2$ denote the negative/positive side of the domain wall of track 1/2. Each one of the four partial conductances is defined as~\cite{LawZBP,FlensbergZBP}:
\begin{align}
\frac{dI_a}{dV}=\frac{2e^2}{h}\frac{\Gamma_a^2}{\Gamma_a^2+(eV)^2}\,,
\end{align}

\noi with the index $a=\{1/2,{\rm N/P}\}$. 

To determine the various $\Gamma_{a}$, we make use of the results of Ref.~\onlinecite{PKSTM} regarding the spin-resolved scanning tunneling spectroscopy of MZMs. After mapping the track to the spin degree of freedom, and given the structure of the eigenvectors in Eq.~\eqref{eq:RacetrackEigen}, we find that the broadening pa\-ra\-me\-ters $\Gamma_{a}$ take the form $\Gamma_{1,{\rm N}}=\Gamma_{2,{\rm P}}=\Gamma\cos^2(\theta/2)$ and $\Gamma_{2,{\rm N}}=\Gamma_{1,{\rm P}}=\Gamma\sin^2(\theta/2)$, with $\Gamma$ the broadening obtained when the respective MZM fully belongs to a single track. Hence, we end up with the following expression for the quadrupolar dif\-fe\-ren\-tial conductance:
\begin{align}
\frac{dI_Q}{dV}=\frac{e^2}{h}\frac{{\cal V}^2\cos\theta}{\left[{\cal V}^2+\cos^4\left(\frac{\theta}{2}\right)\right]\left[{\cal V}^2+\sin^4\left(\frac{\theta}{2}\right)\right]},
\end{align}

\noi where we set ${\cal V}=eV/\Gamma$. During the track exchange, $\theta$ is varied adia\-ba\-ti\-cal\-ly in the interval $\theta=[0,\pi]$. In order to measure the above quantity, it is strategic to convolute the currents with a sinusoidal reference signal of the form $\cos\theta$, and carry out the measurement for a number of $N_c$ cycles in order to cancel out any possible noise contributions. The coefficient of the $\cos\theta$ component of the time-averaged quadrupolar conductance is given by:
\bea
\frac{d\bar{I}_Q^{\cos\theta}}{dV}
&=&\int_0^{2\pi N_c}\frac{2\cos\theta d\theta}{2\pi N_c}\frac{dI_Q}{dV}\no\\
&=&\frac{2e^2}{h}\sqrt{2{\cal V}}\left(2\sqrt[\leftroot{-2}\uproot{3}4]{1+{\cal V}^2}\frac{\cos\zeta}{1/\sqrt{2}}
-\frac{\sqrt{2}\sin\zeta}{\sqrt[\leftroot{-2}\uproot{3}4]{1+{\cal V}^2}}\right)
\quad\phd\eea

\noi where $\zeta=\big(\pi/2+\tan^{-1}{\cal V}\big)/2$. As shown in Fig.~\ref{fig:Figure6}(b) the above component of the averaged quadrupolar differential conductance features a characteristic scaling relation. In fact, the normalized quantity $\big(d\bar{I}_Q^{\cos\theta}/dV\big)/\sqrt{2eV/\Gamma}$ becomes quantized and equal to $2e^2/h$ at $V=0$, similar to the single-lead dif\-fe\-ren\-tial conductance $dI/dV$.

Concluding this paragraph, we point out that the above conclusions hold for a generic two-track system, including double-nanowire hybrids~\cite{Kanne,Vekris,VekrisFullShell}, as long as one can controllably transfer the spectral weight of each MZM from one track/nanowire to the other. 

\section{MZM Fusion - Experimental Knobs and Signatures}\label{sec:VIII}

The above conclusions rely on the fact that the MZMs of the pair remain decoupled while carrying out the spectroscopic measurements.  Nonetheless, signatures of the presence of the MZM pair can be detected by means of their fusion which is achieved by controllably coupling them. Since for the domain wall assumed in Sec.~\ref{sec:IV} both MZM eigenvectors constitute eigenstates of $\tau_1$, the most general coupling Hamiltonian takes the form:
\begin{align}
{\cal H}_{\rm MZM}^{\rm fusion}(\hat{p}_x,x)=\bm{\lambda}\cdot\big[\bm{V}(\hat{p}_x)+\bm{V}'(\hat{p}_x)\tau_1\big]
\end{align}

\noi and violates the chiral symmetry $\Pi=\tau_1$ of Eq.~\eqref{eq:RaceTrackDW}. We introduced $\bm{V}(\hat{p}_x)=\big(V_1\hat{p}_x,V_2,V_3\hat{p}_x\big)$ and $\bm{V}'(\hat{p}_x)=\big(V_1'\hat{p}_x,V_2',V_3'\hat{p}_x\big)$. For the spatial structure of the two vector coupling functions, we considered only the lowest-order contributions in terms of $\hat{p}_x$, which are at the same time compatible with the charge-conjugation $\Xi$ that dictates the BdG Hamiltonian. From the above, we find that the matrix elements of the coupling Hamiltonian restricted to the MZM pair subspace are given by the general form: 
\bea
{\cal H}_{\rm MZM}^{\rm fusion}(\theta)=-\bm{F}\cdot\big(\cos\theta,1,\sin\theta\big)i\gamma_{\rm N}\gamma_{\rm P}\,,\label{eq:MZMfusion}
\eea

\noi where $F_s\propto V_s,\, V_s'$ with $s=1,2,3$. The above structure of the MZM coupling Hamiltonian implies that one can infer crucial information regarding the MZM pair by experimentally measuring the conjugate (in the statistical mechanics sense) vector of $\bm{F}$, which is here denoted $\bm{P}(\theta),$ and is defined as:
\bea
\bm{P}(\theta)=-\left<\frac{d{\cal H}_{\rm MZM}^{\rm fusion}(\theta)}{d\bm{F}}\right>\equiv -\frac{dE_{\rm MZM}^{\rm fusion}(\theta)}{d\bm{F}}\,,
\eea

\noi where $E_{\rm MZM}^{\rm fusion}(\theta)$ defines the contribution of the MZM pair to the energy of the system.

First of all, we observe that obtaining the conjugate field of $P_2$ yields information regarding the fermion parity of the domain wall and subsequently of the racetrack, since we find that: 
\bea
P_2=\left<i\gamma_{\rm N}\gamma_{\rm P}\right>=\big<d_0^\dag d_0-d_0d_0^\dag\big>/2
\eea

\noi which takes the values $\pm1/2$, depending on whether the fermionic state which is formed by the MZM pair and gets annihilated by the operator $d_0=(\gamma_{\rm N}+i\gamma_{\rm P})/\sqrt{2}$, is occupied or not. 

The remaining two fields $F_1$ and $F_3$ behave as a two component vector in track space, since they couple in a dipolar fashion to the MZM pair. One can harness the vectorial nature of these fields to define the topological invariant quantity:
\bea
\nu=\int_0^{2\pi}\frac{d\theta}{2\pi}\left[\hat{P}_3(\theta)\frac{d\hat{P}_1(\theta)}{d\theta}-\hat{P}_1(\theta)\frac{d\hat{P}_3(\theta)}{d\theta}\right]
\eea

\noi in terms of the unit conjugate vector:
\bea
(\hat{P}_1,\hat{P}_3)=(P_1,P_3)/\sqrt{P_1^2+P_3^2}\,.\no
\eea

\noi As described by the above formula, by varying $\theta\in[0,2\pi)$, i.e., by performing a double MZM track exchange, one expects to experimentally observe the vorticity value $\nu=1$. The exact quantization of $\nu$ reflects the presence of the MZM, and is only accessible as long as fermion parity is preserved during the entire process. Hence, any deviations from the quantized value may indicate possible fermion parity switchings and quasiparticle poisoning of the system. 

\begin{table}[t!]
\begin{center}
\begin{tabular}{|c|c|}\hline
\phd\bt{Field}$^{\phantom{|^|}}$&\bt{Physical Quantities}\\\hline
\phd $F_1$$^{\phantom{|^|}}$&$\tilde{M}_{||}\sin\big[\big(\phi_1-\phi_2\big)/2\big]$, $\tilde{M}_{||}\sin\big[\big(\omega_1-\omega_2\big)/2\big]$$^{\phantom{|^|}}$\\\hline
\phd $F_2$$^{\phantom{|^|}}$&$\Phi_z$,\,\,\,$\sin\big[\big(\phi_1-\phi_2\big)/2\big]\sin\big[\big(\omega_1-\omega_2\big)/2\big]$$^{\phantom{|^|}}$\\\hline
\phd $F_3$$^{\phantom{|^|}}$&$J_1-J_2$,\,\,\,$\upsilon_1\sin\big(\phi_1-\omega_1\big)-\upsilon_2\sin\big(\phi_2-\omega_2\big)$$^{\phantom{|^|}}$\\\hline
\end{tabular}
\caption{Physical quantities contributing to the fusion terms appearing in the MZM coupling Hamiltonian of Eq.~\eqref{eq:MZMfusion}.}
\label{Table:FusionTerms}
\end{center}
\end{table}

So far, we have not made any specific mentioning regarding the nature of the physical quantities which can couple the MZMs at the domain wall. To identify sui\-table coupling quantities, we start from the most general Hamiltonian described in Eq.~\eqref{eq:RaceTrack}, and project onto the eigenstates $\eta_1=\sigma_1=1$. This process is described in more detail in App.~\ref{app:AppendixIII}. We identify various terms which contribute to the vector $\bm{F}$, that we present in Table~\ref{Table:FusionTerms}.

As we confirm from Table~\ref{Table:FusionTerms}, there exist various knobs that can be employed to fuse the MZM pair. For exam\-ple, the fermion parity of the pair can be probed by subjec\-ting the MZM pair to the presence of an out-of-plane flux $\Phi_z$. Notably, a recent work~\cite{AbiagueSpinMagnetization} discussed the possibility of inferring the topological properties of planar Rashba-Josephson junction~\cite{Hell,PientkaPlanar,MohantaSkyrmion,Tonio,Ren} by studying the inplane spin suscepti\-bi\-li\-ty. While also in our work the magnetic and magnetization fields lie in the plane, we instead propose that signatures of MZMs can be tracked by measuring the out-of-plane orbital component of the magnetization. In fact, while it is well established that the orbital magnetization encodes information regarding the topological properties of a generic CI~\cite{Niu}, it is of a particular importance in the present case, since there are no out-of-plane magnetic fields applied to the system. Hence, the MZM racetracks proposed here, provide a fertile ground for mapping out the fermion parity through disentangling the MZM pair contribution to the response to out-of-plane fluxes. 

Another option to detect the fermion parity is to induce a superconducting difference $\propto\phi_1-\phi_2$ in the two tracks, accompanied by a magnetic stripe spin-orientation misalignment $\propto\omega_1-\omega_2$. As a consequence, one can either experimentally measure the $4\pi$-periodic Josephson current:
\bea 
J_c\propto \cos\big[\big(\phi_1-\phi_2\big)/2\big]\sin\big[\big(\omega_1-\omega_2\big)/2\big]\no
\eea

\noi generated under the presence of the abovementioned spin-orientation misalignment~\cite{OjanenME,KotetesJ,AliceaJ,PientkaJ}, or, detect a spin current:
\bea J_x\propto \sin\big[\big(\phi_1-\phi_2\big)/2\big]\cos\big[\big(\omega_1-\omega_2\big)/2\big]\no\eea 

\noi flowing from one CI to the other with spin orientation in the $x$ axis, by imposing a phase difference $\phi_1-\phi_2$~\cite{KotetesJ,AliceaJ,PientkaJ,Gilbert}.

We now discuss dipolar bias fields which allow expe\-ri\-men\-tal\-ly probing the underlying twofold ground state degeneracy induced by a MZM pair. From Table~\ref{Table:FusionTerms}, we observe that either a superconducting phase difference $\phi_1-\phi_2$ or magnetic stripe spin-orientation misalignment $\propto\omega_1-\omega_2$ is sufficient to generate a $F_1$ field, as long as $\tilde{M}_{||}$ is nonzero. On the other hand, the $F_3$ field can be either engineered by inducing a difference between the net momenta $J_{1,2}$ that can flow through each MZM track, or, by introducing a mismatch between the mixed phase differences $\phi_1-\omega_1$ and $\phi_2-\omega_2$. In fact, the latter can be alternatively achieved by impo\-sing $\omega_1=\omega_2\neq\phi_1=\phi_2$ as long as the velocities satisfy $\upsilon_1\neq\upsilon_2$. By means of the experimental measurement of the conjugate vector $(P_1,P_3)$ of $(F_1,F_3)$ during a double MZM track-exchange process one infers the emergence of topological order through the possible observation of the topological invariant quantity $\nu$ which constitutes a winding number reflecting the two-component vectorial nature of $(F_1,F_3)$.

\section{Summary and Conclusions}\label{sec:IX}

In this work we expose the first route to employ a charge Chern insulator (CI) for the engineering of a topological superconductor (TSC). Specifically, we demonstrate that pyramidal heterostructures of superconductors (SC) and CIs enable the induction of effective 1D p-wave superconductivity and hence the realization of Majorana-zero-mode (MZM) tracks. A SC/CI/SC where the two SCs are kept at a superconducting phase dif\-fe\-ren\-ce constitutes the fundamental building block of such an architecture. 

MZM braiding becomes accessible in racetracks constructed by SC/CI/SC/CI/SC hybrids where the three SCs are kept at different phases. Up to date, TSCs induced by superconducting phases differences have been mainly discussed in connection to systems dictated by Rashba-like spin-orbit coupling (SOC), such as to\-po\-lo\-gi\-cal insulators~\cite{FuKane,Tanaka}, planar Josephson junctions~\cite{Hell,PientkaPlanar}, and semiconducting nanowires~\cite{KotetesClassi,NoZeeman,Melo,Oreg3Phase}. Remarkably, the MZM racetracks proposed in this work do not rely on the presence of Rashba SOC. Instead, the charge CI is assumed to be spin degenerate and the requirement of antisymmetric SOC is provided by externally imposing a homogeneous magnetic field in conjunction with a transversely oriented magnetic stripe. The magnetization induced by the stripe is required to have an opposite orientation near the two edges. Based on recent theoretical predictions~\cite{FlensbergMag,KlinovajaGraphene,Fatin,ZhouZutic,Abiague,FPTA}, we conclude that engi\-neering such an inhomogeneous magnetization profile is feasible with the currently existing technologies in nanomagnetics~\cite{Kontos,FrolovMag}. The above aspects highlight the enhanced tunability of the MZM racetrack proposed, since the topological properties can be controlled and induced by a variety of external knobs.

We remark that the mechanism underlying the conversion of the charge CI into a TSC crucially relies on a local Andreev reflection mechanism. In more detail, superconductivity is induced on a given edge by virtue of the spin-degenerate nature of the chiral edge modes. Therefore, our proposal is not applicable to the quantum anomalous Hall insulator (QAHI)~\cite{QAHI,LawQuasi1DQAHI}, since there, the pre\-sen\-ce of Rashba SOC leads to spin polarized chiral edge modes. We also note that in contrast to Ref.~\onlinecite{LawQuasi1DQAHI}, here the nontrivial topology relies solely on the chiral edge modes of the CI, and the emergence of a 2D TSC is not a pre\-re\-qui\-si\-te. Even more, our mechanism appears more ge\-ne\-ral and less restrictive than the one discussed in Ref.~\onlinecite{PabloSanJose} for engineering MZMs on a graphene edge in proximity to a conventional SC. There, it is the presence of the two valleys that mediates the local Andreev mechanism and the superconducting proximity effect. In stark contrast, our proposal does not require a multiband structure for the chiral edge modes, since we reside on coupling two opposite edges. Hence, we expect our results to be applicable beyond the charge CIs discussed here and thus to hold for generic quantum Hall systems with (near) spin-degenerate chiral edge modes. Going back to our original motivation, the possible discovery of chiral edge modes in Kagome SCs in the insulating regime promises to open perspectives for realizing such SC/CI/SC based on only Kagome materials residing in the CI and SC phases, or, on Kagome CIs interfaced with other conventional SCs.

We further elaborate on the manipulation and expe\-ri\-men\-tal pro\-bes of MZM pairs. We show that MZM braiding can take place in racetracks construed by SC/CI/SC/CI/SC structures, and relies on the adiabatic control of the intertrack electron tunneling and the spatial profile of the various superconducting phase dif\-fe\-ren\-ces. Our braiding protocol in MZM racetracks relies on the MZM track exchange, which is a process that allows pairs of MZMs to simultaneously swap track. In contrast to prior works which have discussed the spatial exchange of a MZM pair by means of T-junctions~\cite{AliceaTQC}, Y-junctions~\cite{Clarke}, or skyrmion-based racetracks~\cite{Silas}, here the braiding transformation stems solely from the MZM track exchange. As we show in this work, additional processes which shuttle the MZMs in the racetrack in order to complete the MZM exchange, do not contribute to the non-Abelian part of the Berry phase picked up by the twofold-degenerate ground state of the system.

Motivated by the particular geometric features of the MZM platforms discussed here, we further propose sui\-ta\-ble routes to experimentally probe the MZM track exchange. For detecting the former, we propose to measure the time-averaged quadrupolar differential conductance $dI_Q/dV$ which is detectable using four normal leads contacting the CIs at the four positions where MZM are located during the MZM track exchange. The purpose of time averaging is dual. First of all, it is strategic to average over a number of MZM track exchange cycles in order to suppress the impact of the various noise sources that may influence the experimental measurement process. But most importantly, we find that in the clean case, the time-averaged and suitably rescaled $dI_Q/dV$ becomes quantized and equal to a single unit of conductance for $V=0$. Therefore, the present MZM platform allows for an additional quantized spectroscopic quantity apart from the standard single-lead differential conductance. In fact, these two quantities satisfy a scaling relation which can provide a clearer signature of the underlying MZMs. We note that our results have a generic character and may find application in other racetrack type of platforms, these including the double-nanowires setups experimentally realized recently~\cite{Kanne,Vekris,VekrisFullShell}. 

The last component of this work focuses on the experimental control and detection of the fusion of MZM pairs in racetracks. We unveil that a MZM pair can couple to various external fields which can be employed to probe the fermion parity of the pair and in turn the racetrack, as well as to pin down the twofold de\-ge\-ne\-ra\-cy of the ground state. The former becomes possible by experimentally studying the response to flux piercing the cross-sections of the two CIs of the racetrack, or, by monitoring a Josephson/spin current flowing from between the two tracks as a result of imposing combined phase differences and magnetic stripe orientation misalignments. On the other hand, the emergence of non-Abelian topological order can be probed by a two-component vector external field which couples in a dipolar fashion to the MZM pair, due to the 2D spatial distribution of the latter. By extracting the vectorial response to this external vector field, one can construct a winding number which is a topological invariant and when it becomes equal to $\pm1$ reflects the presence of the twofold degenerate ground state and the underlying topological order.    

All in all, our work provides a holistic approach to MZM racetracks using previously not-discussed approaches and material components. We hope that are novel mechanism and experimental considerations will motivate further expe\-ri\-ments in intrinsic Kagome SCs as well as artificial TSCs, and inspire new theoretical concepts concerning the manipulation of MZMs.

\section*{Acknowledments} J.~A.~W. and P.~K. acknowledge funding from the National Na\-tu\-ral Science Foundation of China (Grant No.~12074392). During the year 2020, J.~A.~W. and P.~K. received funding from the project ``Topological Quantum Hall Hybrids'' supported by the CAS Key Laboratory of Theoretical Physics, Institute of Theo\-re\-ti\-cal Physics.

\appendix

\section{Topological Classification Details for the Single-Track Model}\label{app:AppendixI}

In this appendix, we perform the topological classification of the Hamiltonian in Eq.~\eqref{eq:effective_model_1}. When all pa\-ra\-me\-ters are nonzero and $\delta\phi\neq\pi$, the Hamiltonian
lies in class D where only the charge-conjugation symmetry effected by the ope\-ra\-tor $\Xi=\tau_1{\cal K}$ is preserved. For $\tilde{M}_{||}=0$ and $\delta\phi\neq\pi$, Eq.~\eqref{eq:effective_model_1} belongs to class BDI since it additionally possesses time-reversal and chiral symmetries, correspondingly effected by the ope\-ra\-tors $\Theta=\eta_1{\cal K}$ and $\Pi=\tau_1\eta_1$. The topological invariant of a BDI class Hamiltonian in 1D is characterized by a winding number here-termed $w$. Nevertheless, if gap closings only occur at high-symmetry points, class BDI can be also charac\-te\-ri\-zed by the Majorana number ${\cal M}$ which classifies class D Hamiltonians. As we prove below, ${\cal M}$ is indeed here sufficient to map out the topological phase dia\-gram in both BDI and D classes. To show this, we infer $w$ for the case $\tilde{M}_{||}=0$. We employ the unitary transformation generated by the matrix: $\hat{\cal U}=(\tau_1\eta_1+\tau_3)/\sqrt{2}$ to block off-diagonalize the Hamiltonian of Eq.~\eqref{eq:effective_model_1} as follows:
\bea
&&\hat{\cal U}^\dag\hat{\cal H}_{\rm track}(k_x)\hat{\cal U}=-\alpha k_x\tau_2\eta_2+t\tau_1+\mu\tau_1\eta_1+\tilde{M}_\perp\tau_2\eta_2\sigma_2\no\\
&&-B\tau_1\eta_1\sigma_1+\Delta\cos(\delta\phi/2)\tau_2\sigma_2+ \tilde{\Delta}\sin(\delta\phi/2)\tau_1\eta_3\sigma_2\,,\qquad
\eea

\noi where $\tau_\pm=\big(\tau_1\pm i\tau_2\big)/2$. From the above we find that the upper off-diagonal block reads as:
$\hat{A}(k_x)=i\alpha k_x\eta_2+t-i\tilde{M}_\perp\eta_2\sigma_2-B\eta_1\sigma_1-i\Delta\cos(\delta\phi/2)\sigma_2+\tilde{\Delta}\sin(\delta\phi/2)\eta_3\sigma_2+\mu\eta_1$. The winding number is defined as the winding number of $\det[\hat{A}(k_x)]$. Hence, topological phase transitions take place when $\det[\hat{A}(k_x)]=0$. Since, ${\rm Im}\det[\hat{A}(k_x)]=8\alpha k_x\tilde{M}_\perp t\Delta\cos(\delta\phi/2)$, we conclude that within the interval $(-k_c,k_c)$ considered here, the bulk gap closes only at $k_x=0$. Therefore, in order to infer the topological phase diagram it suffices to examine when ${\rm Re}\det[\hat{A}(k_x=0)]=0$. The sign of ${\rm Re}\det[\hat{A}(k_x=0)]=0$ coincides with ${\cal M}$, thus proving that the latter can characterize the to\-po\-lo\-gi\-cal character of the system in both D and BDI cases.

\begin{table}[t!]
\begin{centering}
\begin{tabular}{|c|c|c|c|c|}
\hline 
$\tilde{M}_{||}^{\phantom{||}}=0$&$\mu=0$&$\delta\phi=\pi$&Class&Fully-Gapped Spectrum\\ \hline 
$\times$&$\times$&$\times$&D&
$\times$ ($\checkmark$ only for small $\tilde{M}_{||}^{\phantom{|}}$)\\\hline 
$\times$&$\checkmark$&$\times$&D&
$\times$ ($\checkmark$ only for small $\tilde{M}_{||}^{\phantom{|}}$)\\\hline 
$\times$&$\times$&$\checkmark$&D&
$\times$\\\hline 
$\times$&$\checkmark$&$\checkmark$&A$\oplus$A&
$\times$\\\hline 
$\checkmark$&$\times$&$\times$&BDI&$\checkmark$\\\hline 
$\checkmark$&$\checkmark$&$\times$&BDI&$\checkmark$\\\hline 
$\checkmark$&$\times$&$\checkmark$&BDI&$\checkmark$\\\hline 
$\checkmark$&$\checkmark$&$\checkmark$&AI$\oplus$AI&$\times$\\\hline
\end{tabular}
\par\end{centering}
\caption{Symmetry classification table for the Hamiltonian of Eq.~\eqref{eq:effective_model_1} for particular parameter values. The symbol $\checkmark$ ($\times$) implies that the statement of the top row is true (false).}
\label{Table:class_classification}
\end{table}

We additionally note that for $\tilde{M}_{||}=\mu=0$ and $\delta\phi=\pi$, an additional time-reversal symmetry emerges with $\Theta'=\tau_1\eta_2\sigma_2{\cal K}$. Its presence alters the symmetry class to AI$\oplus$AI which is trivial in 1D and cannot protect MZMs. We also remark that $\tilde{M}_{||}$ is here considered suitably small, in order for the bulk energy spectrum of Eq~\eqref{eq:effective_model_1} to be fully-gapped. However, when $\delta\phi=\pi$, the Hamiltonian of Eq.~\eqref{eq:effective_model_1} leads to gapless trivial phases which cannot support MZMs. We summarize our results in Table~\ref{Table:class_classification}.

{\color{black}

\section{Single Track - Numerical Simulations on the Lattice}\label{app:AppendixII}

The predictions of our low-energy model are verified by considering the full model in the corresponding parameter regime. For our nu\-me\-rics, we employ a lattice extension of our continuum model in Sec.~\ref{sec:II}, which is obtained by performing the mappings $k_{x,y}\mapsto\sin k_{x,y}$ and $k^2\mapsto 2-2(\cos k_x+\cos k_y)$. Hence, the respective lattice extension for the CI reads in $\bm{k}$-space as:
\bea
\hat{H}_{\rm CI;BdG}^{\rm lattice}&=&\frac{1}{2}\sum_{\bm{k}}\bm{\psi}_{\bm{k}}^\dag\big[\alpha\sin k_x\rho_1+\beta\sin k_y\tau_3\rho_2\no\\
&+&\big(2-2\cos k_x-2\cos k_y-m_0\big)\tau_3\rho_3\big]\bm{\psi}_{\bm{k}}\,.\quad\label{eq:latticeCI}
\eea

\noi To obtain the energy spectrum of the system under open boundary conditions in the $x$ axis, we employ the plane wave basis operators $\bm{\psi}_{k_x,R_y}=\nicefrac{1}{\sqrt{W}}\sum_{k_y}e^{iR_yk_y}\bm{\psi}_{\bm{k}}$, with $R_y\in\mathbb{Z}$ the site index in the $y$ axis. We further take into account the Hamiltonian terms in Eqs.~\eqref{eq:Hmag} and~\eqref{eq:pairing}, but with the coordinates properly extended to the lattice. To describe the above terms we additionally extend our formalism to the respective Nambu space in analogy to the formalism adopted in Sec.~\ref{sec:III}.

\begin{figure}[t!]
\begin{center}
\includegraphics[width=0.95\columnwidth]{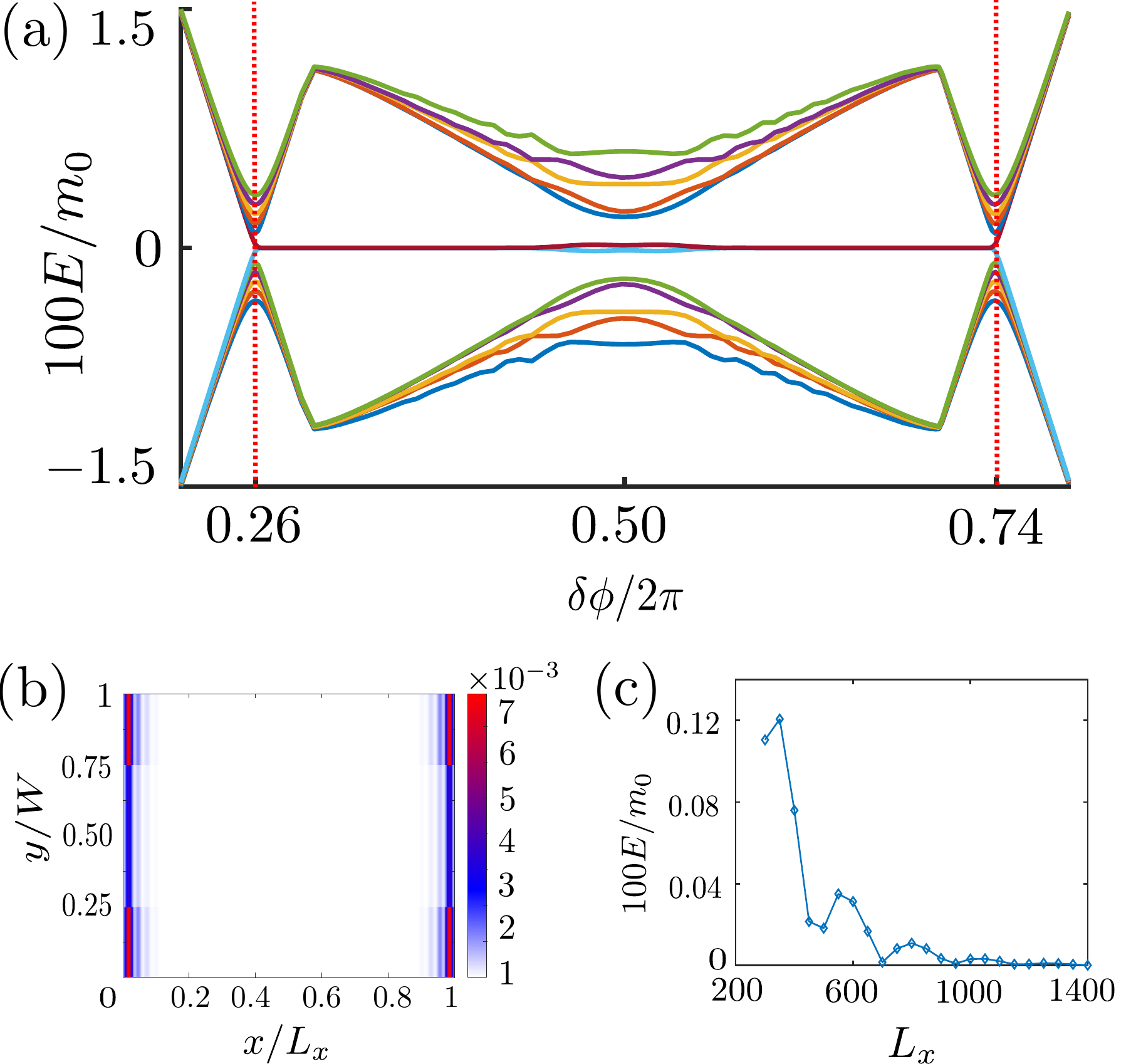}
\end{center}
\caption{{\color{black}(a) Numerically-obtained low-energy spectrum for the lattice version of the model defined by  Eqs.~\eqref{eq:Hmag},~\eqref{eq:pairing}, and~\eqref{eq:latticeCI} as a function of the superconducting phase dif\-ference $\delta\phi$. Based on the results of Eq.~\eqref{eq:MajoranaNumber} for the parameter va\-lues $B=\Delta=0.1$, $M_\perp=0.02$, $M_{||}=\mu=0$, and $W=4$ (leading to $t\approx0.139$, $\tilde{M}_\perp\approx0.020$, and $\tilde{\Delta}\approx0.097$), MZMs are expected to appear for $\delta\phi/2\pi\in(0.26,0.5)\cup(0.5,0.74)$. Indeed this is verified for most of the $\delta\phi$ interval. (b) Spatial weight (squared absolute value of the wavefunction) of each MZM for $\delta\phi/2\pi=0.35$. The MZMs appear on opposite termination edges of the strip. (c) depicts the length-dependence of the lowest energy eigenstate of the system for $\delta\phi/2\pi=0.45$. We find that the nonzero energy Andreev modes observed in (a) near $\delta\phi=\pi$ are a result of finite-size effects, since their ener\-gy decreases upon increasing $L_x$. Even more, the low-energy states for $\delta\phi=\pi$ are not genuine MZMs due to a symmetry class transition taking place for this phase difference value.}}
\label{fig:FigureAppI}
\end{figure}

In our numerics, we investigate a CI strip with a finite length and width, of $L_x=1000$ and $W=4$ sites, respectively. With no loss of generality we assume for simplicity that $|\Delta(R_y)|=\Delta=0.1$ and replace the $\cos\big[Q(y-y_0)\big]$ dependence of the magnetic stripe by a stepwise spatial profile $\bm{M}(R_y)=M_\perp\big[1-2\Theta(R_y-W/2)\big]\hat{\bm{y}}$, with $M_\perp=0.02$. A similar stepwise profile is considered for the superconducting phase $\phi(R_y)=\delta\phi\big[1-2\Theta(R_y-W/2)\big]/2$. We also set $\alpha=\beta=m_0=1$, $B=0.72$, and $\mu=0$. For the above parameter values, the topological phase diagram shown in Fig.~\ref{fig:Figure2} that we obtained using the low-energy Hamiltonian of Eq.~\eqref{eq:effective_model_1}, implies that the hybrid system enters the topologically nontrivial phase for $\delta\phi/2\pi\in(0.26,0.5)\cup(0.5,0.74)$ and two MZMs are expected to appear at the termination edges of the strip.

Indeed, this is verified from our numerical investigations. The energy spectrum resulting from a particular numerical simulation as a function of $\delta\phi$ is shown in Fig.~\ref{fig:FigureAppI}(a). MZMs appear to be stabilized in an extended region which is mainly in accordance with our earlier theoretical predictions. In Fig.~\ref{fig:FigureAppI}(b), we depict the spatial weights of the respective MZM wavefunctions for $\delta\phi/2\pi=0.35$. In agreement with the result of Fig.~\ref{fig:Figure2}, the MZMs evolve into nonzero energy Andreev edge modes for $\delta\phi=\pi$. As explained earlier, this is attributable to the  BDI$\rightarrow$AI$\oplus$AI symmetry class transition which takes place at this special value, and renders the system trivial. Remarkably, we find that MZMs hybridize into nonzero ener\-gy Andreev modes for an extended region centered at $\delta\phi=\pi$. However, this superficial discre\-pan\-cy compared to the analytical low-energy predictions is due to the reduction of resolution in our numerics. Indeed, by repeating our numerical investigations for values near $\delta\phi=\pi$ with strip lengths much larger than $L_x=1000$, we have verified that the energy of the Andreev modes tends to zero and that MZMs are present also in this window. See for instance Fig.~\ref{fig:FigureAppI}(c) which was obtained for $\delta\phi/2\pi=0.45$. 

}

\section{Mappings to p-wave Superconductors} \label{app:AppendixIII}

We here detail the procedure that allows us to map the single- and coupled double-track low-energy Hamil\-to\-nians to single- and coupled double-track p-wave superconductor models. In the upcoming derivation we assume that $t_{1,2}>B_{1,2}>0$ and $\mu_{1,2}>0$. Under the above hie\-rar\-chy and sign conventions, the mapping is performed by projecting the original model Hamil\-to\-nians onto the $\eta_1=\sigma_1=1$ sector. Our analysis follows Ref.~\onlinecite{AliceaTQC} closely.

\begin{widetext} 

\subsection{Mapping for the Single-Track Model}

Our starting point is the Hamiltonian of Eq.~\eqref{eq:effective_model_1}. We proceed by re-expressing this Hamiltonian in the basis of the $\eta_1=\pm1$ eigenstates. By explicitly writing down the matrix elements in this two-state Hilbert subspace we find:
\begin{align}
\hat{\cal H}_{\rm track}^{\eta_1\,{\rm basis}}(k_x)=
\left(\begin{array}{cc}
\tau_3\big(t-\mu-B\sigma_1\big)-\Delta\cos\big(\delta\phi/2\big)\tau_2\sigma_2&
\alpha k_x+\tilde{M}_{||}\tau_3\sigma_1+\tilde{M}_\perp\sigma_2-\tilde{\Delta}\sin\big(\delta\phi/2\big)\tau_1\sigma_2\\
\alpha k_x+\tilde{M}_{||}\tau_3\sigma_1+\tilde{M}_\perp\sigma_2-\tilde{\Delta}\sin\big(\delta\phi/2\big)\tau_1\sigma_2&
-\tau_3\big(t+\mu+B\sigma_1\big)-\Delta\cos\big(\delta\phi/2\big)\tau_2\sigma_2
\end{array}\right)\,.
\end{align}

\noi By assuming that $\tau_3\big(t-\mu-B\sigma_1\big)-\Delta\cos\big(\delta\phi/2\big)\tau_2\sigma_2\approx0$ we find that $-\tau_3\big(t+\mu+B\sigma_1\big)-\Delta\cos\big(\delta\phi/2\big)\tau_2\sigma_2\approx-2t\tau_3$. Therefore, by restricting to energy values $E\approx0$ and by considering $\alpha k_x$, $\tilde{M}$ and $\tilde{\Delta}$ compared to the $2t$ energy splitting between the $\eta_1=\pm1$ states, we find that the effective Hamiltonian dictating the $\eta_1=1$ level reads as:
\bea
\hat{\cal H}_{\rm track}^{\eta_1=1}(k_x)&\approx&
% \tau_3\big(t-\mu-B\sigma_1\big)-\Delta\cos\big(\delta\phi/2\big)\tau_2\sigma_2\qquad\no\\
% &+&\big[\alpha k_x+\tilde{M}_{||}\tau_3\sigma_1+\tilde{M}_\perp\sigma_2-\tilde{\Delta}\sin\big(\delta\phi/2\big)\tau_1\sigma_2\big]\frac{\tau_3}{2t}
% \big[\alpha k_x+\tilde{M}_{||}\tau_3\sigma_1+\tilde{M}_\perp\sigma_2-\tilde{\Delta}\sin\big(\delta\phi/2\big)\tau_1\sigma_2\big]\no\\
% %
% %
% &=&
\tau_3\big(\tilde{t}-\mu-B\sigma_1\big)-\Delta\cos\big(\delta\phi/2\big)\tau_2\sigma_2+\alpha k_x\big(\tilde{M}_{||}\sigma_1+\tilde{M}_\perp\tau_3\sigma_2\big)/t\,,
\eea

\noi where we introduced the renormalized hybridization ener\-gy scale $\tilde{t}=t+\tilde{M}^2/(2t)-\tilde{\Delta}^2\sin^2\big(\delta\phi/2\big)/(2t)$ to fa\-ci\-li\-ta\-te the notation. We now proceed by re-expressing the above in the basis of the $\sigma_1=\pm1$ eigenstates. We have:
\begin{align}
\hat{\cal H}_{\rm track}^{\eta_1=1;\sigma_1=\pm1\,{\rm basis}}(k_x)=
\left(\begin{array}{cc}
\big(\tilde{t}-\mu-B\big)\tau_3+\alpha\tilde{M}_{||} k_x/t
&-i\big[\Delta\cos\big(\delta\phi/2\big)\tau_2-\alpha\tilde{M}_\perp k_x\tau_3/t\big]\\
i\big[\Delta\cos\big(\delta\phi/2\big)\tau_2-\alpha\tilde{M}_\perp k_x\tau_3/t\big]&\big(\tilde{t}-\mu+B\big)\tau_3-\alpha\tilde{M}_{||} k_x/t
\end{array}\right)\,.
\end{align}

\noi Following the lines of the earlier projection, we assume that $\big(\tilde{t}-\mu-B\big)\tau_3+\alpha\tilde{M}_{||} k_x/t\approx0$ which implies that $\big(\tilde{t}-\mu+B\big)\tau_3-\alpha\tilde{M}_{||} k_x/t\approx 2B\tau_3+2\alpha\tilde{M}_{||} k_x/t\approx 2B\tau_3$. Assuming that $\tilde{M}_\perp$ and $\Delta\cos\big(\delta\phi/2\big)$ are smaller than the splitting for $k_x=0$ which is relevant here, we find that the effective Hamiltonian for the $\eta_1=\sigma_1=1$ eigenstate:
\begin{align}
\hat{\cal H}_{\rm track}(k_x)\approx
\big(\tilde{t}-\mu-B\big)\tau_3+\frac{\alpha\tilde{M}_{||}k_x}{t}
-\big[\Delta\cos\big(\delta\phi/2\big)\tau_2-\alpha\tilde{M}_\perp k_x\tau_3/t\big]\frac{\tau_3}{2B}\big[\Delta\cos\big(\delta\phi/2\big)\tau_2-\alpha\tilde{M}_\perp k_x\tau_3/t\big]\,.
\end{align}

\noi Straightforward manipulations provide the expression in Eq.~\eqref{eq:SingleTrack}, after first having projected the unitary matrix in Eq.~\eqref{eq:unitaryTrafo} onto the $\sigma_1=1$ eigenstate.

\subsection{Mapping for the Double-Track Model}

We now repeat the procedure of the paragraph above for two coupled tracks. Our starting point is now the Hamiltonian in Eq.~\eqref{eq:RaceTrack}. In the following, we consider that the two tracks are characterized by generally dif\-fe\-rent phases $\phi_{1,2}$ and $\omega_{1,2}$ but are otherwise identical. Following closely the approach for projecting the Hamiltonian onto the $\eta_1=1$ eigenstate discussed above, we once again assume that $\tau_3\big(t-\mu-B\sigma_1\big)-\Delta\cos\big(\delta\phi/2\big)\tau_2\sigma_2\approx0$ we find that $-\tau_3\big(t+\mu+B\sigma_1\big)-\Delta\cos\big(\delta\phi/2\big)\tau_2\sigma_2\approx-2t\tau_3$. This procedure yields the effective Hamiltonian for $\eta_1=1$:
\bea
&&\hat{\cal H}_{\rm racetrack}^{\eta_1=1}(k_x)
\approx
% \tau_3\big(t-\mu-B\sigma_1\big)-\Delta\cos\big(\delta\phi/2\big)\tau_2\sigma_2+\big(T_S+T_+\big)e^{i\lambda_3\tau_3[(\omega_1-\omega_2)\sigma_1-(\phi_1-\phi_2)]/2}\lambda_1\tau_3\no\\
% &&\qquad\qquad\qquad+\big[\alpha k_x+\tilde{M}_{||}\tau_3\sigma_1+\tilde{M}_\perp\sigma_2-\tilde{\Delta}\sin\big(\delta\phi/2\big)\tau_1\sigma_2+T_-e^{i\lambda_3\tau_3[(\omega_1-\omega_2)\sigma_1-(\phi_1-\phi_2)]/2}i\lambda_2\tau_3\big]\frac{\tau_3}{2t}\no\\
% &&\qquad\ph\phd\qquad\qquad\big[\alpha k_x+\tilde{M}_{||}\tau_3\sigma_1+\tilde{M}_\perp\sigma_2-\tilde{\Delta}\sin\big(\delta\phi/2\big)\tau_1\sigma_2-T_-e^{i\lambda_3\tau_3[(\omega_1-\omega_2)\sigma_1-(\phi_1-\phi_2)]/2}i\lambda_2\tau_3\big]\no\\
%
%&&=
\tau_3\big[\tilde{t}+T_-^2/(2t)-\mu-B\sigma_1\big]-\Delta\cos\big(\delta\phi/2\big)\tau_2\sigma_2+\frac{\alpha k_x}{t}\big(\tilde{M}_{||}\sigma_1+\tilde{M}_\perp\tau_3\sigma_2\big)+Te^{i\hat{\vartheta}(\sigma_1)\lambda_3\tau_3}\lambda_1\tau_3\no\\
&&\qquad-\frac{T_-}{t}\left[\tilde{M}_\perp\sin\left(\frac{\omega_1-\omega_2}{2}\right)e^{-i\lambda_3\tau_3(\phi_1-\phi_2)/2}\lambda_1\tau_3\sigma_3+\tilde{\Delta}\sin\big(\delta\phi/2\big)\sin\left(\frac{\phi_1-\phi_2}{2}\right)e^{i\lambda_3\tau_3\sigma_1(\omega_1-\omega_2)/2}\lambda_1\tau_2\sigma_2\right]
\eea

\noi where $\hat{\vartheta}(\sigma_1)=\sigma_1(\omega_1-\omega_2)/2-(\phi_1-\phi_2)/2$. The above expression is now projected onto the $\sigma_1=1$ eigenstate under the assumption that: $\tau_3\big(\tilde{t}+T_-^2/(2t)-\mu+B\big)-\tilde{M}_{||}\alpha k_x/t
+Te^{i\hat{\vartheta}(\sigma_1=-1)\lambda_3\tau_3}\lambda_1\tau_3\approx 2B\tau_3$. We now project the Hamiltonian onto the $\sigma_1=1$ eigenstate and conclude with the following expression:
\bea
&&\hat{{\cal H}}_{\rm racetrack}(k_x)\approx
Jk_x+\upsilon k_x\tau_2+m\tau_3+Te^{i\hat{\vartheta}(\sigma_1=1)\lambda_3\tau_3}\lambda_1\tau_3+\frac{T_-\Delta\tilde{\Delta}\sin\big(\delta\phi\big)}{2Bt}\sin\big[(\phi_1-\phi_2)/2\big]e^{i(\omega_1-\omega_2)\lambda_3\tau_3/2}\lambda_1\tau_3\no\\
&&+\big[T_-\tilde{M}_\perp/(Bt)\big]\sin\big[(\phi_1-\phi_2)/2\big]\left\{\Delta\cos\big(\delta\phi/2\big)\sin\big[(\omega_1-\omega_2)/2\big]\lambda_2\tau_1+\tilde{\Delta}\sin\big(\delta\phi/2\big)\cos\big[(\omega_1-\omega_2)/2\big]\frac{\alpha k_x}{t}\lambda_1\tau_2\right\}.\qquad
\eea

\noi Finally, we briefly note that under the following assumption: $\tau_3\big(\tilde{t}+T_-^2/(2t)-\mu+B\big)-\tilde{M}_{||}\alpha k_x/t+Te^{i\hat{\vartheta}(\sigma_1=-1)\lambda_3\tau_3}\lambda_1\tau_3\approx 2B\tau_3-2\tilde{M}_{||}\alpha k_x/t$, we also obtain contributions accompanied by the matrices $\lambda_1$ and $\lambda_1\tau_1$, which become nonzero when either $\tilde{M}_{||}\sin\big[(\phi_1-\phi_2)/2\big]$, or, $\tilde{M}_{||}\sin\big[(\omega_1-\omega_2)/2\big]$ are nonzero.

\end{widetext}

\begin{figure*}[t!]
\begin{centering}
\includegraphics[width=1.0\textwidth]{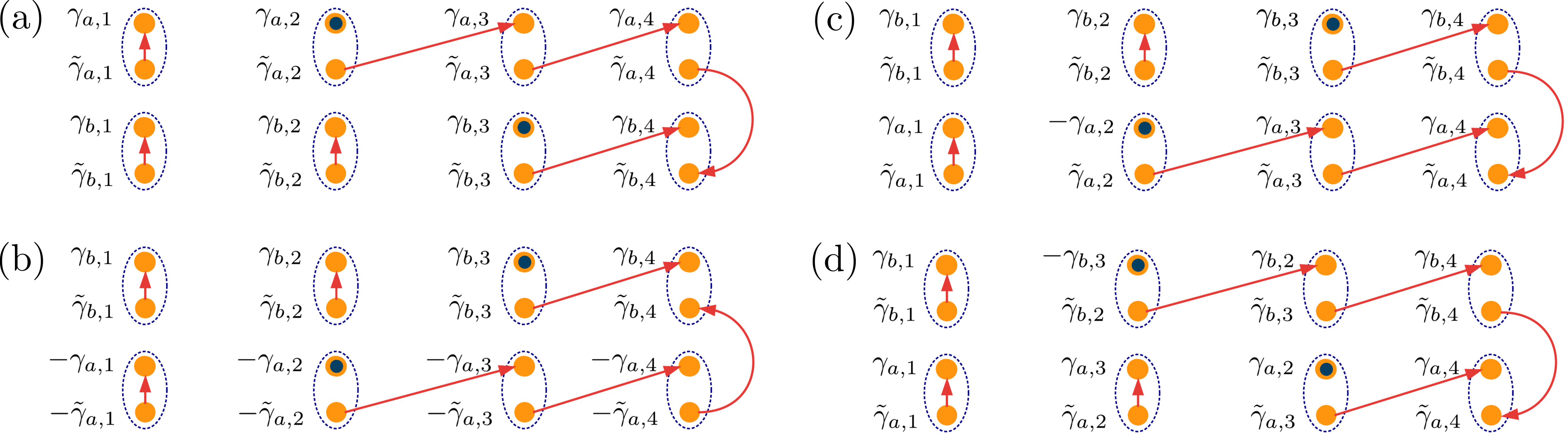}
\par\end{centering}
\caption{MZM braiding using a racetrack consisting of two coupled Kitaev chains supporting unpaired MZMs. (a) shows the configuration before the braiding protocols is initiated. (b) shows the evolved MZMs after the track exchange. Here, the track exchange process swaps the two chains in a rigid manner. The MZMs of one of the two chains pick up a minus sign. (c) is obtained from (b) after removing the braiding unrelated signs. (d) Braiding is complete after shuttling the two MZMs.}
\label{fig:FigureAppII}
\end{figure*}

\section{MZM Braiding Implementation}\label{app:AppendixIV}

In the main text, we discussed how MZM braiding takes place and we emphasized the crucial role of the here-termed MZM track exchange process. In this appendix, we use the above mapping of the MZM racetrack to a racetrack of two coupled p-wave SCs to further elaborate on the details underlying the braiding process. To render the discussion transparent and facilitate an ana\-ly\-ti\-cal treatment we specifically demonstrate how the braiding protocol proposed here is implemented for a topologically-equivalent MZM racetrack consisting of two Kitaev chain models~\cite{KitaevUnpaired}. Even more, we restrict to the sweet spot with unpaired MZMs. The first advantage of examining this special case is that it still allows to generalize results concerning topological properties to more complex Hamiltonian configurations. This is under the condition that the gap of the system remains open when deforming the Kitaev chain model to address other Hamiltonian configurations. The second advantage is that since each MZM has spatial support only on a single lattice site, one can consider chains of arbi\-tra\-ri\-ly short lengths to study the braiding process. In the remainder we study chains of four lattice sites. See also Fig.~\ref{fig:FigureAppII}.

Under the above conditions, a single Kitaev chain consists of two types of MZMs, i.e., $\gamma_n$ and $\tilde{\gamma}_n$, which satisfy: $\{\gamma_n,\gamma_m\}=\{\tilde{\gamma}_n,\tilde{\gamma}_m\}=\delta_{nm}\phd{\rm and}\phd\{\gamma_n,\tilde{\gamma}_m\}=0\phd\forall\ph n,m$, where $n,m$ label the lattice cites. Pairs of $\gamma_n$ and $\tilde{\gamma}_n$ MZMs create an electron $\psi_n=(\gamma_n+i\tilde{\gamma}_n)/\sqrt{2}$. Assu\-ming the sweet spot case, the topologically-trivial segments of a Kitaev chain are dictated by the local coupling of MZMs. See for instance the first sites of either chain $a$ or $b$, in Fig.~\ref{fig:FigureAppII}(a). The red arrows represent Hamiltonian terms of the form $i\tilde{\gamma}_{a,n}\gamma_{a,n}$ and $i\tilde{\gamma}_{b,n}\gamma_{b,n}$, and the direction of the arrow is employed to define the sign of the coupling matrix element. In topologically-nontrivial segments MZMs couple in a nonlocal fashion according to the Hamiltonian terms $i\tilde{\gamma}_{a,n}\gamma_{a,n+1}$ and $i\tilde{\gamma}_{b,n}\gamma_{b,n+1}$. We remark that during the braiding process the two tracks need to remain electronically connected, and here this is achieved by coupling the two unpaired MZMs on the right edge of the two tracks through the term $i\tilde{\gamma}_{a,4}\tilde{\gamma}_{b,4}$.

In Fig.~\ref{fig:FigureAppII} we present the necessary steps to implement braiding in a 4-site Kitaev chain racetrack. The MZMs shown in each panel should be understood within the Heisenberg picture of evolution, and each MZM depicted at a given site in panels (b), (c) and (d) is expressed in terms of the MZMs defined at the beginning of the protocol, i.e., in panel (a). The first necessary step for brai\-ding is the MZM track exchange. Here, we assume that all the sites of the chains $a$ and $b$ become rigidly exchanged. See Fig.~\ref{fig:FigureAppII}(b). Importantly, the MZM wavefunctions of one of the two chains pick up a minus sign due to the structure of the MZM eigenvectors in Eq.~\eqref{eq:RacetrackEigen} and the transformation property in Eq.~\eqref{eq:MZM_Eigenvector_Trafo}. We remark that the sign changes of MZMs which are coupled and belong to the same chain do not affect the brai\-ding outcome. This is because the Hamiltonian remains invariant under such simultaneous sign changes and the extra minus sign can be absorbed by a redefinition of the Fock space states. Changing the sign of one of two MZMs that belong to different chains also does not affect the braiding outcome, since $i\tilde{\gamma}_{a,4}\tilde{\gamma}_{b,4}\mapsto i\tilde{\gamma}_{b,4}(-\tilde{\gamma}_{a,4})=i\tilde{\gamma}_{a,4}\tilde{\gamma}_{b,4}$. 

Figure~\ref{fig:FigureAppII}(c) is obtained from (b) after removing the extra signs of coupled MZMs in the lower chain. The removal of the extra sign in $-\tilde{\gamma}_{a,4}$ is accompanied by the reversal of the coupling arrow. Hence, from panel (c) we find that the two MZMs that need to be braided (depicted with black discs) already carry the extra relative sign. Braiding becomes complete by moving each MZM to the original location of the other MZM. In order to shuttle a MZM, one couples the MZM in question with one nearby already paired-up MZM. For example, in order to shuttle $\gamma_{b,3}$ in Fig.~\ref{fig:FigureAppII}(c) to the left by one site, we need to couple it to $\tilde{\gamma}_{b,2}$ so that $\gamma_{b,3}$ and $\tilde{\gamma}_{b,2}$ become exchanged. At the end of the process, one of the MZMs $\gamma_{b,3}$ or $\tilde{\gamma}_{b,2}$ picks up an extra minus sign~\cite{Clarke}. By employing the same convention for the shuttling process in the two chains, we find that the MZM which we desire to shuttle is the one that picks up the extra sign.
Hence, since the two MZMs that we wish to braid move by an equal amount of sites, their shuttling does not change the relative sign that was picked up at the end of the track exchange.


\begin{thebibliography}{99}

\bibitem{UCDW_KVSb} Y.-X. Jiang, J.-X. Yin, M. M. Denner, N. Shumiya, B. R. Ortiz, G. Xu, Z. Guguchia, J. He, Md S. Hossain, X. Liu, J. Ruff, L. Kautzsch, S. S. Zhang, G. Chang, I. Belopolski, Q. Zhang, T. A. Cochran, D. Multer, M. Litskevich, Z.-J. Cheng, X. P. Yang, Z. Wang, R. Thomale, T. Neupert, S. D. Wilson, and M. Zahid Hasan, \textit{Unconventional chiral charge order in kagome superconductor KV$_3$Sb$_5$}, Nat. Mater. (2021). https://doi.org/10.1038/s41563-021-01034-y. 

\bibitem{muonGraf} E. M. Kenney, B. R. Ortiz, C. Wang, S. D. Wilson,
and M. J. Graf, \textit{Absence of local moments in the
kagome metal KV$_3$Sb$_5$ as determined by muon spin spectroscopy},  J. Phys.: Condens. Matter \bt{33}, 235801 (2021).

\bibitem{muonKVSb} C. Mielke III, D. Das, J.-X. Yin, H. Liu, R. Gupta, C. N. Wang, Y.-X. Jiang, M. Medarde, X. Wu, H. C. Lei, J. J. Chang, P. Dai, Q. Si, H. Miao, R. Thomale, T. Neupert, Y. Shi, R. Khasanov, M. Z. Hasan, H. Luetkens, Z. Guguchia, \textit{Time-reversal symmetry-breaking charge order in a correlated kagome superconductor}, arXiv:2106.13443.

\bibitem{OpticalDetecCDW} E. Uykur, B. R. Ortiz, S. D. Wilson, M. Dressel, and A. A. Tsirlin, \textit{Optical detection of charge-density-wave instability in the non-magnetic kagome metal KV$_3$Sb$_5$}, arXiv:2103.07912.

\bibitem{NematicCdwKVSb} H. Li, H. Zhao, B. R. Ortiz, T. Park, M. Ye, L. Balents, Z. Wang, S. D. Wilson, and I. Zeljkovic, \textit{Rotation symmetry breaking in the normal state of a kagome superconductor KV$_3$Sb$_5$}, arXiv:2104.08209.

\bibitem{XrayEPcoupling} H. Luo, Q. Gao, H. Liu, Y. Gu, D. Wu, C. Yi, J. Jia, S. Wu, X. Luo, Y. Xu, L. Zhao, Q. Wang, H. Mao, G. Liu, Z. Zhu, Y. Shi, K. Jiang, J. Hu, Z. Xu, and X. J. Zhou, \textit{Electronic Nature of Charge Density Wave and Electron-Phonon Coupling in Kagome Superconductor KV$_3$Sb$_5$}, arXiv:2107.02688. 

\bibitem{AHE_CDW_CVSb} F. H. Yu, T. Wu, Z. Y. Wang, B. Lei, W. Z. Zhuo, J. J. Ying, and X. H. Chen, \textit{Concurrence of anomalous Hall effect and charge density wave in a superconducting topological kagome metal}, Phys. Rev. B \bt{104}, L041103 (2021).

\bibitem{CascadeofPhasesCVSb} H. Zhao, H. Li, B. R. Ortiz, S. M. L. Teicher, T. Park, M. Ye, Z. Wang, L. Balents, S. D. Wilson, and I. Zeljkovic, \textit{Cascade of correlated electron states in a kagome superconductor CsV$_3$Sb$_5$}, arXiv:2103.03118.

\bibitem{CDWvortexCVSb} Z. Liang, X. Hou, F. Zhang, W. Ma, P. Wu, Z. Zhang, F. Yu, J.-J. Ying, K. Jiang, L. Shan, Z. Wang, and X.-H. Chen, \textit{Three-dimensional charge density wave and robust zero-bias conductance peak inside the superconducting vortex core of a kagome superconductor CsV$_3$Sb$_5$}, Phys. Rev. X \bt{11}, 031026 (2021).

\bibitem{FSmappingCVSb} B. R. Ortiz, S. M. L. Teicher, L. Kautzsch, P. M. Sarte, N. Ratcliffe, J. Harter, J. P. C. Ruff, R. Seshadri, and S. D. Wilson, \textit{Fermi surface mapping and the nature of charge density wave order in the kagome superconductor CsV$_3$Sb$_5$}, arXiv:2104.07230v2.

\bibitem{HiddenFluxPhaseCVSb} L. Yu, C. Wang, Y. Zhang, M. Sander, S. Ni, Z. Lu, S. Ma, Z. Wang, Z. Zhao, H. Chen, K. Jiang, Y. Zhang, H. Yang, F. Zhou, X. Dong, S. L. Johnson, M. J. Graf, J. Hu, H.-J. Gao, and Z. Zhao, \textit{Evidence of a hidden flux phase in the topological kagome metal CsV$_3$Sb$_5$}, arXiv:2107.10714.

\bibitem{UCDW_RbVSb} N. Shumiya, M. S. Hossain, J.-X. Yin, Y.-X. Jiang, B. R. Ortiz, H. Liu, Y. Shi, Q. Yin, H. Lei, S. S. Zhang, G. Chang, Q. Zhang, T. A. Cochran, D. Multer, M. Litskevich, Z.-J. Cheng, X. P. Yang, Z. Guguchia, S. D. Wilson, and M. Zahid Hasan, \textit{Intrinsic nature of chiral charge order in the kagome superconductor RbV$_3$Sb$_5$}, Phys. Rev. B \bt{104}, 035131 (2021).

\bibitem{BinghaiCDW} H. Tan, Y. Liu, Z. Wang, and B. Yan, \textit{Charge density waves and electronic properties of superconducting kagome metals}, Phys. Rev. Lett. \bt{127}, 046401 (2021).

\bibitem{ChiralFluxKagome} X. Feng, K. Jiang, Z. Wang, and J. Hu, \textit{Chiral flux phase in the Kagome superconductor AV$_3$Sb$_5$}, Science Bulletin \bt{66}, 1384 (2021).

\bibitem{Titus} M. M. Denner, R. Thomale, and T. Neupert, \textit{Analysis of charge order in the kagome metal AV$_3$Sb$_5$ (A=K,Rb,Cs)}, arXiv:2103.14045.

\bibitem{Nandkishore} Y.-P. Lin and R. M. Nandkishore, \textit{Complex charge density waves at Van Hove singularity on hexagonal lattices: Haldane-model phase diagram and potential realization in kagome metals AV$_3$Sb$_5$}, Phys. Rev. B \bt{104}, 045122 (2021).

\bibitem{KunJiangClassification} X. Feng, Y. Zhang, K. Jiang, J. Hu, \textit{Low-energy effective theory and symmetry classification of flux phases on Kagome lattice}, arXiv:2106.04395.

\bibitem{BinghaiCDWGeometry} H. Miao, H. X. Li, H. N. Lee, A. Said, H. C. Lei, J. X. Yin, M. Z. Hasan, Z. Wang, H. Tan, and B. Yan, \textit{Geometry of the charge density wave in kagom\'e metal AV$_3$Sb$_5$}, arXiv:2106.10150.

\bibitem{CDWChristensen} M. H. Christensen, T. Birol, B. M. Andersen, and R. M. Fernandes, \textit{Theory of the charge-density wave in AV$_3$Sb$_5$ kagome metals}, arXiv:2107.04546.

\bibitem{Venderbos} J. W. F. Venderbos, \textit{Symmetry analysis of translational symmetry broken density waves: Application to hexagonal lattices in two dimensions}, Phys. Rev. B \bt{93}, 115107 (2016).

\bibitem{KagomeRichInvHS} Y. Hu, X. Wu, B. R. Ortiz, S. Ju, X. Han, J. Z. Ma, N. C. Plumb, M. Radovic, R. Thomale, S. D. Wilson, A. P. Schnyder, and M. Shi, \textit{Rich Nature of Van Hove Singularities in Kagome Superconductor CsV$_3$Sb$_5$}, arXiv:2106.05922.

\bibitem{Haldane} F. D. M. Haldane, \textit{Model for a quantum Hall effect without Landau levels: condensed-matter realization of the parity anomaly}, Phys. Rev. Lett. \bt{61}, 2015 (1988).

\bibitem{VarmaPRB97} C. M. Varma, \textit{Non-Fermi-liquid states and pairing instability of a general model of copper oxide metals},  Phys. Rev. B \bt{55}, 14554 (1997).

\bibitem{VarmaPseudogap} C. M. Varma, \textit{Theory of the pseudogap state of the cuprates}, Phys. Rev. B \bt{73}, 155113 (2006).

\bibitem{VarmaJCMP} C. M. Varma, \textit{Pseudogap in cuprates in the loop-current ordered state}, J. Phys. Condens. Matter \bt{26}, 505701 (2014).

\bibitem{NayakUDWs} C. Nayak, \textit{Density-wave states of nonzero angular momentum}, Phys. Rev. B \bt{62}, 4880 (2000).

\bibitem{ChakravartyHiddenOrder} S. Chakravarty, R. B. Laughlin, D. K. Morr, and C. Nayak, \textit{Hidden order in the cuprates}, Phys. Rev. B \bt{63}, 094503 (2001).

\bibitem{Yakovenko90} V. M. Yakovenko, \textit{Chern-Simons Terms and n Field in Haldane's Model for the Quantum Hall Effect without Landau Levels}, Phys. Rev. Lett. \bt{65}, 251 (1990).

\bibitem{Tewari} S. Tewari, C. Zhang, V. M. Yakovenko, and S. Das
Sarma, Phys. Rev. Lett. \bt{100}, 217004 (2008).

\bibitem{KotetesEPL} P. Kotetes and G. Varelogiannis, \textit{Spontaneous Quantum Hall Effect in chiral d-density waves}, EPL \bt{84}, 37012 (2008).

\bibitem{KotetesPRBR} P. Kotetes, G. Varelogiannis, \textit{Meissner effect without superconductivity from a chiral d-density wave}, Phys. Rev. B \bt{78}, 220509(R) (2008).

\bibitem{ChuanweiZhang} C. Zhang, S. Tewari, V. M. Yakovenko, and S. Das Sarma, \textit{Anomalous Nernst effect from a chiral d-density wave state in underdoped cuprate superconductors}, Phys. Rev. B \bt{78}, 174508 (2008).

\bibitem{KotetesPRL} P. Kotetes and G. Varelogiannis, \textit{Chirality Induced Tilted-Hill Giant Nernst Signal}, Phys. Rev. Lett. \bt{104}, 106404 (2010).

\bibitem{QAHIprop1} C.-X. Liu, X.-L. Qi, X. Dai, Z. Fang, and S.-C. Zhang, \textit{Quantum Anomalous Hall Effect in Hg$_{1-y}$Mn$_y$Te Quantum Wells}, Phys. Rev. Lett. \bt{101}, 146802 (2008).

\bibitem{QAHIprop2} R. Yu, W. Zhang, H. J. Zhang, S. C. Zhang, X. Dai, and Z. Fang, \textit{Quantized Anomalous Hall Effect in Magnetic Topological Insulators}, Science \bt{329}, 61 (2010).

\bibitem{QAHIexp} C.-Z. Chang, J. Zhang, X. Feng, J. Shen, Z. Zhang, M. Guo, K. Li, Y. Ou, P. Wei, L.-L. Wang, Z.-Q. Ji, Y. Feng, S. Ji, X. Chen, J. Jia, X. Dai, Z. Fang, S.-C. Zhang, K. He, Y. Wang, L. Lu, X.-C. Ma, and Q.-K. Xue, \textit{Experimental Observation of the Quantum Anomalous Hall Effect in a Magnetic Topological Insulator}, Science \bt{340}, 167 (2013).

\bibitem{QAHI} X.-L. Qi, T. L. Hughes, and S.-C. Zhang, \textit{Chiral Topological Superconductor From the Quantum Hall State}, Phys. Rev. B \bt{82}, 184516 (2010).

\bibitem{LawQuasi1DQAHI} C.-Z. Chen, Y.-M. Xie, J. Liu, P. A. Lee, and K. T. Law, \textit{Quasi-one-dimensional Quantum Anomalous Hall Systems as New Platforms for Scalable Topological Quantum Computation}, Phys. Rev. B \bt{97}, 104504 (2018).

\bibitem{CMM} Q.-L. He, L. Pan, A. L. Stern, E. Burks, Xiaoyu Che, G. Yin, J. Wang, B. Lian, Q. Zhou, E. S. Choi, K. Murata, X. Kou, T. Nie, Q. Shao, Y. Fan, S.-C. Zhang, K. Liu, J. Xia, and K. L. Wang, \textit{Chiral Majorana edge state in a quantum anomalous Hall insulator-superconductor structure}, Science \bt{357}, 294 (2017).

\bibitem{CMMM} J. Shen, J. Lyu, J. Z. Gao, Y. -M. Xie, C.-Z. Chen, C.-W. Cho, O. Atanov, Z. Chen, K. Liu, Y. J. Hu, K. Y. Yip, S. K. Goh, Q. L. He, L. Pan, K. L. Wang, K. T. Law, and R. Lortz, \textit{Spectroscopic Fingerprint of Chiral Majorana Modes at the Edge of a Quantum Anomalous Hall Insulator/Superconductor Heterostructure}, PNAS \bt{117}, 238 (2019).

\bibitem{WenCMM} W. Ji and X.-G. Wen, \textit{A mechanism of $\frac{1}{2}\frac{e^2}{h}$ conductance plateau without 1D chiral Majorana fermions},  Phys. Rev. Lett. \bt{120}, 107002 (2018).

\bibitem{SauCMM} Y. Huang, F. Setiawan, and J. D. Sau, \textit{Disorder-induced half-integer quantized conductance plateau in quantum anomalous Hall insulator-superconductor structures}, Phys. Rev. B \bt{97}, 100501 (2018).

\bibitem{NoCMM} M. Kayyalha, D. Xiao, R. Zhang, J. Shin, J. Jiang, F. Wang, Y.-F. Zhao, L. Zhang, K. M. Fijalkowski, P. Mandal, M. Winnerlein, C. Gould, Q. Li, L. W. Molenkamp, M. H. W. Chan, N. Samarth, and C.-Z. Chang, \textit{Non-Majorana Origin of the Half-Quantized Conductance Plateau in Quantum Anomalous Hall Insulator and Superconductor Hybrid Structures}, Science \bt{367}, 64 (2020). 

{\color{black}

\bibitem{Rodenbach} L. K. Rodenbach, I. T. Rosen, E. J. Fox, P. Zhang, L. Pan, K. L. Wang, M. A. Kastner, and D. Goldhaber-Gordon, \textit{Bulk dissipation in the quantum anomalous Hall effect}, APL Mater. \bt{9}, 081116 (2021).

\bibitem{Ferguson} G. M. Ferguson, R. Xiao, A. R. Richardella, D. Low, N. Samarth, and K. C. Nowack, \textit{Direct visualization of electronic transport in a quantum anomalous Hall insulator}, arXiv:2112.13122 (2021).

\bibitem{Rosen} Ilan T. Rosen, Molly P. Andersen, Linsey K. Rodenbach, Lixuan Tai, Peng Zhang, Kang L. Wang, M. A. Kastner, David Goldhaber-Gordon, \textit{Measured potential profile in a quantum anomalous Hall system suggests bulk-dominated current flow}, arXiv:2112.13123 (2021).

}

\bibitem{BiaoLian} B. Lian, X.-Q. Sun, A. Vaezi, X.-L. Qi, and S.-C. Zhang, \textit{Topological Quantum Computation Based on Chiral Majorana Fermions}, PNAS \bt{115}, 10938 (2018). 

{\color{black}

\bibitem{KitaevTQC} A. Y. Kitaev, \textit{Fault-Tolerant Quantum Computation by Anyons}, Ann. Phys. \bt{303}, 2 (2003).

\bibitem{NayakTQC} C. Nayak, S. H. Simon, A. Stern, M. Freedman, and S. Das Sarma, \textit{Non-Abelian Anyons and Topological Quantum Computation}, Rev. Mod. Phys. \bt{80}, 1083 (2008).

}

\bibitem{KotetesClassi} P. Kotetes, \textit{Classification of Engineered 
Topological Superconductors}, New J. Phys. \bt{15}, 105027 (2013). 

\bibitem{Heimes} A. Heimes, P. Kotetes, and G. Sch\"on, \textit{Majorana fermions from Shiba states in an antiferromagnetic chain on top of a superconductor}, Phys. Rev. B \bt{90}, 060507(R) (2014).

\bibitem{PabloSanJose} P. San-Jose, J. L. Lado, R. Aguado, F. Guinea, and J. Fern\'andez-Rossier, \textit{Majorana Zero Modes in Graphene}, Phys. Rev. X \bt{5}, 041042 (2015).

\bibitem{Livanas} G. Livanas, M. Sigrist, and G. Varelogiannis, \textit{Alternative paths to realize Majorana Fermions in Superconductor-Ferromagnet Heterostructures}, Sci. Rep. \bt{9}, 6259 (2019).

{\color{black}

\bibitem{Kontos} M. M. Desjardins, L. C. Contamin, M. R. Delbecq, M. C. Dartiailh, L. E. Bruhat, T. Cubaynes, J. J. Viennot, F.
Mallet, S. Rohart, A. Thiaville, A. Cottet, and T. Kontos,
Synthetic spin orbit interaction for Majorana devices, Nat. Mater. \bt{18}, 1060 (2019).

\bibitem{FrolovMag} M. J. A. Jardine, J. P. T. Stenger, Y. Jiang, E. J. de Jong, W. Wang, A. C. Bleszynski Jayich, and S. M. Frolov, \textit{Integrating micromagnets and hybrid nanowires for topological quantum computing}, arXiv:2104.05130. 

}

\bibitem{MarraMParticle} P. Marra, D. Inotani, and M. Nitta, \textit{Dispersive 1D Majorana modes with emergent supersymmetry in 1D proximitized superconductors via spatially-modulated potentials and magnetic fields}, arXiv:2106.09047.

\bibitem{SauProxi} J. D. Sau, R. M. Lutchyn, S. Tewari, and S. Das Sarma, \textit{Robustness of Majorana fermions in proximity-induced superconductors}, Phys. Rev. B \bt{82}, 094522 (2010).

\bibitem{Potter} A. C. Potter and P. A. Lee, \textit{Engineering a p+ip superconductor: Comparison of topological insulator and Rashba spin-orbit-coupled materials}, Phys. Rev. B \bt{83}, 184520 (2011).

\bibitem{AZ} A. Altland and M. R. Zirnbauer, \textit{Nonstandard Symmetry Classes in Mesoscopic Normal-Superconducting Hybrid Structures}, Phys. Rev. B \bt{55}, 1142 (1997).

\bibitem{SchnyderClassi} A. P. Schnyder, S. Ryu, A. Furusaki, and A. W. W. Ludwig, \textit{Classification of topological insulators and superconductors in three spatial dimensions}, Phys. Rev. B \bt{78}, 195125 (2008). 
 
\bibitem{Vuik} A. Vuik, D. Eeltink, A. R. Akhmerov, and M. Wimmer,
\textit{Effects of the electrostatic environment on the Majorana
nanowire devices}, New J. Phys. \bt{18}, 033013 (2016).

\bibitem{AntipovPRX} A. E. Antipov, A. Bargerbos, G. W. Winkler, B. Bauer, E. Rossi, and R. M. Lutchyn, \textit{Effects of Gate-Induced Electric Fields on Semiconductor Majorana Nanowires}, Phys. Rev. X \bt{8}, 031041 (2018).

\bibitem{WoodsSP} B. D. Woods, T. D. Stanescu, and S. Das Sarma, \textit{Effective theory approach to the Schr\"odinger-Poisson problem in semiconductor Majorana devices}, Phys. Rev. B \bt{98}, 035428 (2018).

\bibitem{MikkelsenPRX} A. E. G. Mikkelsen, P. Kotetes, P. Krogstrup, and K. Flensberg, \textit{Hybridization at Superconductor-Semiconductor Interfaces}, Phys. Rev. X \bt{8}, 031040 (2018).

\bibitem{Reeg} C. Reeg, D. Loss, and J. Klinovaja, \textit{Metallization of Rashba wire by superconducting layer in the strong-proximity regime}, Phys. Rev. B \bt{97}, 165425 (2018).

\bibitem{KitaevUnpaired} A. Y. Kitaev, \textit{Unpaired Majorana Fermions in Quantum Wires}, Phys. Usp. \bt{44}, 131 (2001).

\bibitem{OjanenME} T. Ojanen, \textit{Magnetoelectric effects in superconducting nanowires with Rashba spin-orbit coupling}, Phys. Rev. Lett. \bt{109}, 226804 (2012).

\bibitem{Sticlet2013} D. Sticlet, C. Bena, and P. Simon, \textit{Josephson Effect in Superconducting Wires Supporting Multiple Majorana Edge States}, Phys. Rev. B \bt{87}, 104509 (2013).

\bibitem{PKsynthetic} P. Kotetes, M. T. Mercaldo, and M. Cuoco, \textit{Synthetic Weyl Points and Chiral Anomaly in Majorana Devices with Nonstandard Andreev-Bound-State Spectra}, Phys. Rev. Lett. \bt{123}, 126802 (2019).

\bibitem{MTMPRB} M. T. Mercaldo, P. Kotetes, and M. Cuoco, \textit{Magnetoelectrically-Tunable Andreev-Bound-State Spectra and Spin Polarization in P-Wave Josephson Junctions}, Phys. Rev. B \bt{100}, 104519 (2019).

\bibitem{Riwar} R.-P. Riwar, M. Houzet, J. S. Meyer, and Y. V. Nazarov, \textit{Multi-Terminal Josephson Junctions as Topological Matter}, Nat. Commun. \bt{7}, 11167 (2016).

\bibitem{Eriksson} E. Eriksson, R.-P. Riwar, M. Houzet, J. S. Meyer, and Y. V. Nazarov, \textit{Topological transconductance quantization in a four-terminal Josephson junction}, Phys. Rev. B \bt{95}, 075417 (2017).

\bibitem{Meyer_PRL} J. S. Meyer and Manuel Houzet, \textit{Nontrivial Chern Numbers in Three-Terminal Josephson Junctions}, Phys. Rev. Lett. \bt{119}, 136807 (2017).

\bibitem{LevchenkoI} H.-Y. Xie, M. G. Vavilov, and  A. Levchenko, \textit{Topological Andreev bands in three-terminal Josephson junctions}, Phys. Rev. B \bt{96}, 161406 (2017).

\bibitem{LevchenkoII} H.-Y. Xie, M. G. Vavilov, and A. Levchenko, \textit{Weyl nodes in Andreev spectra of multiterminal Josephson junctions: Chern numbers, conductances, and supercurrents}, Phys. Rev. B \bt{97}, 035443 (2018).

\bibitem{Belzig} H. Weisbrich, R. L. Klees, G. Rastelli, and W. Belzig, \textit{Second Chern Number and Non-Abelian Berry Phase in Topological Superconducting Systems}, PRX Quantum \bt{2}, 010310 (2021).  

\bibitem{Rastelli} R. L. Klees, J. C. Cuevas, W. Belzig, and G. Rastelli, \textit{Many-body Quantum Geometry in Superconductor-Quantum Dot Chains}, Phys. Rev. B \bt{103}, 014516 (2021)  

\bibitem{WeylCircuits} V. Fatemi, A. R. Akhmerov, and L. Bretheau, \textit{Weyl Josephson Circuits}, Phys. Rev. Research \bt{3}, 013288 (2021).

\bibitem{LevchenkoNonAbelian} H.-Y. Xie, Jaglul Hasan, and A. Levchenko, \textit{Non-Abelian monopoles in the multiterminal Josephson effect}, arXiv:2107.03435.

\bibitem{Mi} B. van Heck, S. Mi, and A. R. Akhmerov, \textit{Single fermion manipulation via superconducting phase differences in multiterminal Josephson junctions}, Phys. Rev. B \bt{90}, 155450 (2014).

\bibitem{Balseiro} L. P. Gavensky, G. Usaj, and C. A. Balseiro, \textit{Topological phase diagram of a three-terminal Josephson junction: From the conventional to the Majorana regime}, Phys. Rev. B \bt{100}, 014514 (2019). 

\bibitem{Sakurai} K. Sakurai, M. T. Mercaldo, S. Kobayashi, A. Yama\-ka\-ge, S. Ikegaya, T. Habe, P. Kotetes, M. Cuoco, and Y. Asano, \textit{Nodal Andreev Spectra in Multi-Majorana Three-Terminal Josephson Junctions}, Phys. Rev. B \bt{101}, 174506 (2020).

\bibitem{Houzet} J. S. Meyer and M. Houzet, \textit{Conductance quantization in topological Josephson trijunctions}, Phys. Rev. B \bt{103}, 174504 (2021).

\bibitem{Draelos} A. W. Draelos, M.-T. Wei, A. Seredinski, H. Li, Y. Mehta, K. Watanabe, T. Taniguchi, I. V. Borzenets, F. Amet, and G. Finkelstein, \textit{Supercurrent Flow in Multiterminal Graphene Josephson Junctions}, Nano Lett. \bt{19}, 1039 (2019).

\bibitem{Manucharyan} N.~Pankratova, H.~Lee, R.~Kuzmin, M.~ Vavilov, K.~Wickramasinghe, W.~Mayer, J.~Yuan, J.~ Shabani, and V.~E.~Manucharyan, \emph{The multi-terminal Josephson effect}, Phys. Rev. X \bt{10}, 031051 (2020). 

\bibitem{Arnault}  E. G. Arnault, T. Larson, A. Seredinski, L. Zhao, S. Idris, A. McConnell, K. Watanabe, T. Taniguchi, I. V. Borzenets, F. Amet, and G. Finkelstein, \textit{The Multi-terminal Inverse AC Josephson Effect}, arXiv:2012.15253.  

\bibitem{GrapheneRMP} A. H. Castro Neto, F. Guinea, N. M. R. Peres, K. S. Novoselov, and A. K. Geim, \textit{The electronic properties of graphene}, Rev. Mod. Phys. \bt{81}, 109 (2009).

\bibitem{AliceaTQC} J. Alicea, Y. Oreg, G. Refael, F. von Oppen, and M. P. A. Fisher, \textit{Non-Abelian Statistics and Topological Quantum Information Processing in 1D Wire Networks}, Nat. Phys. \bt{7}, 412 (2011).

\bibitem{SauZBP} J. D. Sau, S. Tewari, R. Lutchyn, T. Stanescu, and S. Das Sarma, \textit{Non-Abelian quantum order in spin-orbit-coupled semiconductors: The search for topological Majorana particles in solid state systems}, Phys. Rev. B \bt{82}, 214509 (2010).

\bibitem{LawZBP} K. T. Law, P. A. Lee, and T. K. Ng, \textit{Majorana Fermion Induced Resonant Andreev Reflection}, Phys. Rev. Lett. \bt{103}, 237001 (2009).

\bibitem{FlensbergZBP} K. Flensberg, \textit{Tunneling characteristics of a chain of Majorana bound states}, Phys. Rev. B \bt{82}, 180516(R) (2010).

\bibitem{PKSTM} P. Kotetes, D. Mendler, A. Heimes, and G. Sch\"on, \textit{Majorana fermion fingerprints in spin-polarised scanning tunneling microscopy}, Physica E: Low-dimensional Systems and Nanostructures \bt{74}, 614 (2015).

\bibitem{Kanne} T. Kanne, D. Olsteins, M. Marnauza, A. Vekris, J. C. Estrada Salda${\rm \tilde{n}}$a, S. Loric, R. D. Schlosser, D. Ross, S. Csonka, K. Grove-Rasmussen, and J. Nyg\aa rd, \textit{Double nanowires for hybrid quantum devices}, arXiv:2103.13938.

\bibitem{Vekris} A. Vekris, J. C. Estrada Salda${\rm \tilde{n}}$a, T. Kanne, M. Marnauza, D. Olsteins, F. Fan, X. Li, T. Hvid-Olsen, X. Qiu, H. Xu, J. Nyg\aa rd, and K. Grove-Rasmussen, \textit{Josephson junctions in double nanowires bridged by in-situ deposited superconductors}, arXiv:2104.01591.

\bibitem{VekrisFullShell} A. Vekris, J. C. Estrada Salda${\rm \tilde{n}}$a, J. de Bruijckere, S. Lori\'c, T. Kanne, M. Marnauza, D. Olsteins, J. Nyg\aa rd, and K. Grove-Rasmussen, \textit{Asymmetric Little-Parks Oscillations in Full Shell Double Nanowires}, arXiv:2106.01181.

\bibitem{AbiagueSpinMagnetization} J. D. Pakizer and A. Matos-Abiague, \textit{Signatures of Topological Transitions in the Spin Susceptibility of Josephson Junctions}, Phys. Rev. B \bt{104}, 100506 (2021).

\bibitem{Hell} M. Hell, M. Leijnse, and K. Flensberg, \textit{Two-dimensional platform for networks of Majorana bound states}, Phys. Rev. Lett. \bt{118}, 107701 (2017).

\bibitem{PientkaPlanar} F. Pientka, A. Keselman, E. Berg, A. Yacoby, A. Stern, and B. I. Halperin, \textit{Topological Superconductivity in a Planar Josephson Junction}, Phys. Rev. X \bt{7}, 021032 (2017).

\bibitem{MohantaSkyrmion} N. Mohanta, S. Okamoto, and E. Dagotto, \textit{Skyrmion Control of Majorana States in Planar Josephson Junctions}, Commun. Phys. \bt{4}, 163 (2021).

\bibitem{Tonio} A. Fornieri, A. M. Whiticar, F. Setiawan, E. Portol\'{e}s Mar\'{i}n, A. C. C. Drachmann, A. Keselman, S. Gronin, C. Thomas, T. Wang, R. Kallaher, G. C. Gardner, E. Berg, M. J. Manfra, A. Stern, C. M. Marcus, and F. Nichele, \textit{Evidence of topological superconductivity in planar Josephson junctions}, Nature \bt{569}, 89 (2019).

\bibitem{Ren} H. Ren, F. Pientka, S. Hart, A. Pierce, M. Kosowsky, L. Lunczer, R. Schlereth, B. Scharf, E. M. Hankiewicz, L. W. Molenkamp, B. I. Halperin, and A. Yacoby, \textit{Topolo\-gi\-cal Superconductivity in a Phase-Controlled Josephson Junction}, Nature \bt{569}, 93 (2019).

\bibitem{Niu} D.~Xiao, M.-C.~Chang, and Q.~Niu, \textit{Berry phase effects on electronic properties}, Rev. Mod. Phys. \bt{82}, 1959 (2010).

\bibitem{KotetesJ} P. Kotetes, A. Shnirman, and G. Sch\"on, \textit{Engineering and Manipulating Topological Qubits in 1D Quantum Wires}, J. Korean Phys. Soc. \bt{62}, 1558 (2013).

\bibitem{AliceaJ} L. Jiang, D. Pekker, J. Alicea, G. Refael, Y. Oreg, A. Brataas, and F. von Oppen, \textit{Magneto-Josephson Effects in Junctions with Majorana Bound States}, Phys. Rev. B \bt{87}, 075438 (2013). 

\bibitem{PientkaJ} F. Pientka, L. Jiang, D. Pekker, J. Alicea, G. Refael, Y. Oreg, and F. von Oppen, \textit{Magneto-Josephson Effects and Majorana Bound States in Quantum Wires}, New J. Phys. \bt{15}, 115001 (2013).

\bibitem{Gilbert} Q. Meng, V. Shivamoggi, T. L. Hughes, M. J. Gilbert, and S. Vishveshwara, \textit{Fractional spin Josephson effect and electrically controlled magnetization in quantum spin Hall edges}, Phys. Rev. B \bt{86}, 165110 (2012).

\bibitem{FuKane} L. Fu and C. L. Kane, \textit{Superconducting Proximity Effect and Majorana Fermions at the Surface of a Topological Insulator}, Phys. Rev. Lett. \bt{100}, 096407 (2008).

\bibitem{Tanaka} Y. Tanaka, T. Yokoyama, and N. Nagaosa, \textit{Manipulation of the Majorana Fermion, Andreev Reflection, and Josephson Current on Topological Insulators}, Phys. Rev. Lett. \bt{103}, 107002 (2009).

\bibitem{NoZeeman} P. Kotetes, \textit{Topological superconductivity in Rashba semiconductors without a Zeeman field}, Phys. Rev. B \bt{92}, 014514 (2015); Erratum Phys. Rev. B \bt{101}, 209904 (2020).

\bibitem{Melo} A. Melo, S. Rubbert, and A. R. Akhmerov, \textit{Supercurrent-induced Majorana bound states in a planar geometry}, SciPost Phys. \bt{7}, 039 (2019).

\bibitem{Oreg3Phase} O. Lesser, K. Flensberg, F. von Oppen, and Y. Oreg, \textit{Three-phase Majorana zero modes at tiny magnetic fields}, Phys. Rev. B \bt{103}, L121116 (2021).

\bibitem{FlensbergMag} M. Kjaergaard, K. W\"olms, and K. Flensberg, \textit{Majorana Fermions in Superconducting Nanowires without Spin-Orbit Coupling}, Phys. Rev. B \bt{85}, 020503(R) (2012).

\bibitem{KlinovajaGraphene} J. Klinovaja and D. Loss, \textit{Giant Spin-Orbit Interaction Due to Rotating Magnetic Fields in Graphene Nanoribbons}, Phys. Rev. X \bt{3}, 011008 (2013).

\bibitem{Fatin} G. L. Fatin, A. Matos-Abiague, B. Scharf, and I. ${\rm \check{Z}}$uti\'c, \textit{Wireless Majorana Bound States: From Magnetic Tunability to Braiding}, Phys. Rev. Lett. \bt{117}, 077002 (2016).

\bibitem{ZhouZutic} T. Zhou, N. Mohanta, J. E. Han, A. Matos-Abiague, and I. ${\rm \check{Z}}$uti\'c, \textit{Tunable magnetic textures in spin valves: From spintronics to Majorana bound states}, Phys. Rev. B \bt{99}, 134505 (2019).

\bibitem{Abiague} N. Mohanta, T. Zhou, J.-W. Xu, J. E. Han, A. D. Kent, J. Shabani, I. ${\rm \check{Z}}$uti\'c, and A. Matos-Abiague, \textit{Electrical Control of Majorana Bound States Using Magnetic Stripes}, Phys. Rev. Applied \bt{12}, 034048 (2019).

\bibitem{FPTA} G.-Y. Huang, B. Li, X.-F. Yi, J.-B. Fu, X. Fu, X.-G. Qiang, P. Xu, J.-J. Wu, C.-L. Yu, P. Kotetes, and M.-T. Deng, \textit{Field-Programmable Topological Array: Framework and Case-Studies}, arXiv:2010.02130.

\bibitem{Clarke} D. J. Clarke, J. D. Sau, and S. Tewari, \textit{Majorana fermion exchange in quasi-one-dimensional networks}, Phys. Rev. B \bt{84}, 035120 (2011).

\bibitem{Silas} S. A. D\'iaz, J. Klinovaja, D. Loss, and S. Hoffman, \textit{Majorana Bound States Induced by Antiferromagnetic Skyrmion Textures}, Phys. Rev. B \bt{104}, 214501 (2021).

\end{thebibliography}
\end{document}